\documentclass[fleqn,usenatbib]{mnras}
\usepackage{newtxtext,newtxmath}
\usepackage[T1]{fontenc}

\DeclareRobustCommand{\VAN}[3]{#2}
\let\VANthebibliography\thebibliography
\def\thebibliography{\DeclareRobustCommand{\VAN}[3]{##3}\VANthebibliography}

\usepackage{graphicx}	% Including figure files
\usepackage{amsmath}	% Advanced maths commands
\usepackage{multirow}
\title{Detection of Quasi-periodic Oscillations in the $\gamma$-Ray Light Curve of 4FGL J0309.9-6058}

\newcommand{\orcid}[1]{\href{https://orcid.org/#1}{\includegraphics[width=10pt]{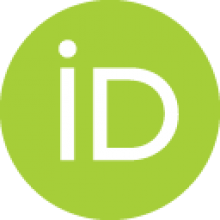}}}

\author[Jingyu Wu]{
Jingyu Wu$^{1}$\orcid{0009-0008-2290-4349},
Zhihao Ouyang$^{1}$\orcid{0009-0000-4102-9115},
Hubing Xiao$^{1}$\thanks{hubing.xiao@shnu.edu.cn}\orcid{0000-0001-8244-1229},
Elisa Prandini$^{2}$,
Shangchun Xie$^{1}$,
Sheng~Yang$^{3}$\orcid{0000-0002-2898-6532},
\newauthor
Jianzhen Chen$^{1}$\thanks{jzchen@shnu.edu.cn},
Shaohua Zhang$^{1}$\orcid{0000-0001-8485-2814},
Haoyang Zhang$^{4}$\orcid{0000-0003-3392-320X},
Junhui Fan$^{5,6,7}$\orcid{0000-0002-5929-0968},
\\
$^{1}$Shanghai Key Lab for Astrophysics, Shanghai Normal University, Shanghai 200234, People's Republic of China\\
$^{2}$INFN, Universit{\`a} di Padova, I-35131 Padova, Italy\\
$^{3}$Institute for Gravitational Wave Astronomy, Henan Academy of Sciences, Zhengzhou 450046, Henan, China\\
$^{4}$Department of Astronomy, Key Laboratory of Astroparticle Physics of Yunnan Province, Yunnan University, Kunming 650091, People’s Republic of China\\
$^{5}$Center for Astrophysics, Guangzhou University, Guangzhou, 510006, People's Republic of China\\
$^{6}$Key Laboratory for Astronomical Observation and Technology of Guangzhou, Guangzhou, 510006, People's Republic of China\\
$^{7}$Astronomy Science and Technology Research Laboratory of Department of Education of Guangdong Province, Guangzhou, 510006, People's Republic of China
}
\date{Accepted XXX. Received YYY; in original form ZZZ}

\pubyear{2025}

\begin{document}
\label{firstpage}
\pagerange{\pageref{firstpage}--\pageref{lastpage}}
\maketitle

\begin{abstract}
In this work, we report, for the first time, a quasi-periodic oscillation (QPO) in the $\gamma$-ray band of 4FGL J0309.9-6058, also known as PKS 0308-611.
We employed three analytical methods (the Lomb-Scargle periodogram, REDFIT, and the weighted wavelet Z-transform) to analyze the QPO signal using \textit{Fermi} $\gamma$-ray light curve data. 
The analysis reveals a potential QPO during MJD 57983$-$60503, with a period of approximately 550 days and a maximum local significance of 3.72$\sigma$ and global significance of 2.72$\sigma$ derived from the WWZ analysis. 
% To further validate this result, we applied Gaussian Process (GP) to the same light curve. 
% The GP model independently confirms a consistent period of $\sim$560 days, further supporting the presence of the QPO.
To validate this result, we applied Gaussian Process (GP) to the same light curve, which independently confirms the presence of QPO signal consistent with our Fourier-based results.
We further extended the analysis to the full duration of the \textit{Fermi} observations, and the results consistently support and strengthen the presence of this QPO signal.
Additionally, a time lag between the optical and $\gamma$-ray bands indicates separate emission regions for these two bands.
Given the year-like timescale of the QPO signal and the fact that a QPO signal with local significance over 3$\sigma$ for full \textit{Fermi}-LAT observed time, we suggest that the QPO is most likely caused by a precessing jet.

\end{abstract}


%%%%%%%%%%%%%%%%%%%%%%%%%%%%%%%%%%%%%%%%%%%%%%%%%%

%%%%%%%%%%%%%%%%% BODY OF PAPER %%%%%%%%%%%%%%%%%%

\section{Introduction}
Active galactic nuclei (AGNs) are luminous central regions of a small fraction of galaxies, emitting more radiation than their host galaxies. 
The extraordinary brightness is due to the energy released as matter within the accretion disk loses angular momentum and gravitational potential energy and falls into the center supermassive black hole (SMBH) \citep{Salpeter1964, Lynden-Bell1969, Urry1995PASP}. 
AGNs can be broadly classified into jet-type and non-jet-type categories based on the presence of their jets \citep{Padovani2017FrASS}.

Blazars represent a distinctive class of AGN, characterized by their radio-loud property and relativistic jets pointed toward the observer. 
This unique orientation results in extreme properties, including rapid variability and strong emission across the entire electromagnetic spectrum, from radio bands to high-energy gamma rays. 
The spectral energy distribution (SED) of blazars typically exhibits a double-peaked structure: the low-energy component, which spans from radio to X-ray wavelengths, is produced by synchrotron radiation of relativistic electrons, while the high-energy component, which extends from X-ray to $\gamma$-ray bands, is generated by inverse Compton scattering of soft photons or through hadronic processes \citep{Blandford1979, Mucke2001, Abdo2010ApJ716, Ghisellini2009MNRAS397, Fan2016ApJS, Wang2022ApJ, Xiao2024RAA24, Ouyang2025ApJ980}.

Quasi-periodic oscillations (QPOs) are used to study the emission mechanisms of blazars. 
They have been detected across a wide range of timescales in different bands.
The most famous example is the BL Lac object OJ 287, which exhibits a QPO in the optical band with a period of approximately 12 years based on over a century of monitoring \citep{Sillanpa1988ApJ, Valtonen2006ApJ, Fan2010RAA}.
In addition, sources such as 1ES 1959+650, 3C 66A, B2 1633+38, 1823+568, 3C 454.3, 3C 273 have been reported QPO signals in the optical band \citep{Schramm1993A&A, Fan2014ApJS, Fan2018AJ, Otero-Santos2020MNRAS, Dong2022RAA, Li2022RAA22}.
Some sources have observed QPO signals in the X-ray and radio bands, such as PKS 0607-157 and 3C 454.3 \citep{Qian2007ChJAA, Li2023RAA23}
The launch of the Large Area Telescope (LAT) aboard the Fermi Gamma-ray Space Telescope in 2008 has significantly enhanced the capability to conduct all-sky monitoring across various time scales \citep{Atwood2009apj} and provided the possibility of discovering QPOs in the $\gamma$-ray band. 
The first QPO source in the $\gamma$-ray band was observed in PG 1553+113, which showed a period of about 2 years \citep{Ackermann2015ApJ}.
Based on over sixteen years of LAT data, more than 30 QPO signals in the $\gamma$-ray band were reported in blazars with periods ranging from months to years \citep[e.g.,][]{Sandrinelli2016, Prokhorov2017MNRAS, Zhang2017ApJ835, Bhatta2019MNRAS, Zhang2021ApJ, Zhang2023PASP}.
There are many physical mechanisms that explain the phenomenon of QPO, such as binary supermassive black hole systems \citep{Lehto1996ApJ, Villata1998MNRAS},
plasma blob helically moving forward along the jet \citep{Camenzind1992A&A}, 
jet precession \citep{Abraham1998ApJ, Abraham1999A&A}, magnetic reconnection events within the jet \citep[e.g.,][]{Huang2013RAA}, and hot spots on the accretion disk revolving around the black hole \citep{Mangalam1993ApJ, Chakrabarti1993ApJ}.

We searched for the QPO signals for each blazar in the \textit{Fermi}-LAT Light Curve Repository (LCR) catalog using the Lomb-Scargle Periodogram (LSP) method and found a significant QPO signal in 4FGL J0309.9-6058.
Thus, in this work, we report a QPO signal of the distant FSRQ object 4FGL J0309.9-6058 ($z=1.479$) in the $\gamma$-ray band.
And in this paper, we use the flat $\Lambda$CDM model with $H_{0}$=67.66 km Mpc$^{-1}$ s$^{-1}$ and $\Omega_{M}$=0.31 \citep{Planck2020A&A}.
This paper is arranged as follows:
We present the observation and data processing in Section 2;
we show the analysis and results of QPO in Section 3;
we provide the discussion and conclusions in Sections 4 and 5.

\section{Fermi-LAT data reduction} 
We collected the LAT data events from the \textit{Fermi}-LAT Pass 8 database 
within a 15$^\circ$ radius region of interest (ROI) centered on 4FGL J0309.9-6058. 
The data spans from MJD 54683
% (August 5, 2008) 
to MJD 60443
% (October 10, 2024) 
and covers an energy range of 0.1 $-$ 300 GeV.
The data analysis was conducted using the latest \texttt{Fermitools} \citep[v2.2.0;][]{Fermi2019asclsoft05011F} and the instrument response functions (IRFs) \texttt{P8R3\_SOURCE\_V3}.
% {\footnote{\url{https://fermi.gsfc.nasa.gov/ssc/data/analysis/documentation/Cicerone/Cicerone\_LAT\_IRFs/IRF\_overview.html}}}. 
The maximum zenith angle value of 90$^\circ$ was selected to avoid the background $\gamma$-rays from the Earth's limb.
The condition ``\texttt{evclass=128, evtype=3}" was used to filter events with a high probability of being photons, and ``\texttt{(DATA\_QUAL$\geqslant$0)\&\&(LAT\_CONFIG==1)}" was used to select the good time intervals. 
The model file, generated by \texttt{make4FGLxml} python package, 
% \footnote{\url{https://github.com/physicsranger/make4FGLxml}},
included all the sources from the Fermi-LAT Fourth Source Catalog (4FGL-DR4; \citealp{Abdollahi2022ApJS}) within 20$^{\circ}$ of the target source, as well as the Galactic (\texttt{gll\_iem\_v07.fits}) and extragalactic isotropic (\texttt{iso\_P8R3\_SOURCE\_V3\_v1.txt}
% {\footnote{\url{https://fermi.gsfc.nasa.gov/ssc/data/access/lat/BackgroundModels.html}}}
) diffuse emission components. 
The normalization parameters and spectral indices of the sources within $5^\circ$ of the target, as well as those of the sources within the region of interest with a variability index (VI) $\geqslant$ 24.725 \citep{Abdollahi2022ApJS}, were set as free parameters.
We checked through the likelihood analysis results, assuming a power-law model, and compared it with a log-parabola model.
The result of test statistic for curve spectrum ${\rm TS_{curve}} = -2(\log \mathcal{L}_{\rm PL}-\log \mathcal{L}_{\rm LP}) < 9$ showed that the log-parabola model is not significantly preferred over the power-law model, where $\mathcal{L}_{\rm PL}$ and $\mathcal{L}_{\rm LP}$ represent the maximum likelihood values obtained from a power-law and a log-parabola fits \citep{Abdollahi2020ApJS}.
Consequently, the target source spectrum is best described by the power-law model, which was used to generate the 30-day binned light curve using the binned likelihood method.
We used the test statistic ${\rm TS} = -2(\log \mathcal{L}_{\rm nosource} - \log \mathcal{L}_{\rm source})$ to calculate the significance of this source ($\mathcal{L}_{\rm source/nosource}$ represents the likelihood of the data given the model with or without a source present at a given position).
We included only flux data points significantly detected with TS $\geqslant$ 9, while the 95\% confidence level upper limit flux values were calculated using the \texttt{UpperLimits}{\footnote{\url{https://fermi.gsfc.nasa.gov/ssc/data/analysis/scitools/upper_limits.html}}}
python tool for cases where TS $< 9$.
The light curve is shown in panel (a) of Figure \ref{fig:QPO_0.1_300}.

\section{ Quasi-periodic Oscillation analysis and results} 
With visual inspection, we found a possible periodic variability during the campaign of MJD 57983$-$60503. 
In order to identify the existence of a QPO signal and to quantify the period, we employed the Lomb-Scargle Periodogram (LSP), REDFIT, and weighted wavelet Z-transform (WWZ).

\subsection{Lomb–Scargle periodogram analysis} 
Lomb-Scargle periodogram \citep[LSP;][]{Lomb1976, Scargle1982} is a widely used method for analyzing periodic signals in time series data. 
The advantage of LSP is that it can handle non-uniformly sampled data, unlike the traditional Fourier transform. 
For irregular sampling, the LSP method iteratively fits sinusoidal curves with different frequencies to light curves and constructs the periodogram according to the goodness of fit, and can provide accurate frequency and power spectrum intensities.
We computed the LSP power using the \texttt{lomb-scargle}\footnote{\url{https://docs.astropy.org/en/stable/timeseries/lombscargle.html}} class provided by \texttt{astropy} and 
setting the frequency range of $f_{\rm min}=\frac{1}{T}$ to $f_{\rm max} = \frac{N}{2 T}$ (which corresponds to the Nyquist frequency $f_{\rm Nyq}$, $T$ represents the total period of observation) with a step of 0.00015, and also considered the flux uncertainties in the analysis.
The LSP power indicates a prominent peak at the timescale of 561.29 days, with its uncertainty estimated from the full width at half maximum (FWHM) of the Gaussian function fitted to the peak, as shown in panel (c) of Figure \ref{fig:QPO_0.1_300}.
To assess the influence of the time sampling on the periodogram, particularly the presence of upper limits in the light curve which were treated as non-detections and excluded from the periodic analysis, we constructed a synthetic light curve by assigning constant flux values to the observed time sampling. 
The LSP of this sampling pattern (the so-called spectral window function) shows no significant peaks at or near the period identified in the analysis, confirming that the detected QPO ($P$=561.29 days) is intrinsic to the source variability rather than an artifact of the sampling pattern.  
We used the false-alarm probability (FAP) to evaluate the confidence level of the LSP peak, whose functional form is as follows: 
$\rm FAP(P_n) = 1-(1-prob(P>P_n))^M$,
where the independent trials $M$ is defined by $M = T \Delta f$ with $\Delta f = f_{\rm Nyq} - f_{\rm min}$. 
The FAP denotes the probability that at least one of the M independent power values in a given frequency band of the white noise periodogram is greater than or equal to the power threshold $\rm P_n$ \citep{Horne1986ApJ}.
\citet{Baluev2008} have given the method ``\texttt{baluev}", which employs extreme value statistics to compute an upper bound of the false alarm probability for the alias-free case.
So we used the ``\texttt{baluev}" method to determine the false-alarm level (FAL) at a 99.99\% FAP, which shows in panel (c) of Figure \ref{fig:QPO_0.1_300} with green line, indicating that there is only a 0.01\% chance of observing such a high peak under the null hypothesis that the data contains no periodic signal.

\subsection{REDFIT}
The light curves of AGNs are mainly affected by red noise, which results from some stochastic processes in a jet plasma or the accretion disk \citep{Li2017ApJ}.
For the non-uniform sampled data, it is difficult to accurately estimate the red-noise spectrum.
REDFIT \citep{Schulz2002CG} was developed to address this issue by directly fitting a first-order autoregressive (AR1) process to unevenly spaced time-series data, thus avoiding interpolation in the time domain and its inherent biases.
As the emission fluxes of AGN are usually autoregressive \citep{Schulz2002CG}, we can use the AR1 process to model the emission red-noise spectrum.
The program \texttt{REDFIT3.8e}\footnote{\url{https://www.marum.de/Prof.-Dr.-michael-schulz/Michael-Schulz-Software.html}} can estimate the spectrum using LSP and Welch overlapped segment averaging (WOSA).
We set the oversampling factor for LSP (\texttt{ofac}) to 10, the number of WOSA segments ($n_{50} = 1$), and selected the Welch spectral window to reduce spectral leakage.
The REDFIT provides a maximum significance level at FAP of 99\% corresponding to confidence levels of 2.58$\sigma$, which is estimated from the power spectrum against the red-noise background in the AR1 process \citep{Schulz2002CG}. 
As shown in panel (d) of Figure \ref{fig:QPO_0.1_300}, a distinct peak emerges at a timescale of 548.16 days with a significance level exceeding 99\%. 
The periodicity uncertainty is estimated from the FWHM of the Gaussian function fitted to the REDFIT peak.

\subsection{Weighted wavelet Z-transform analysis} 
\begin{figure*}
\centering
\includegraphics[width=6.9 in]{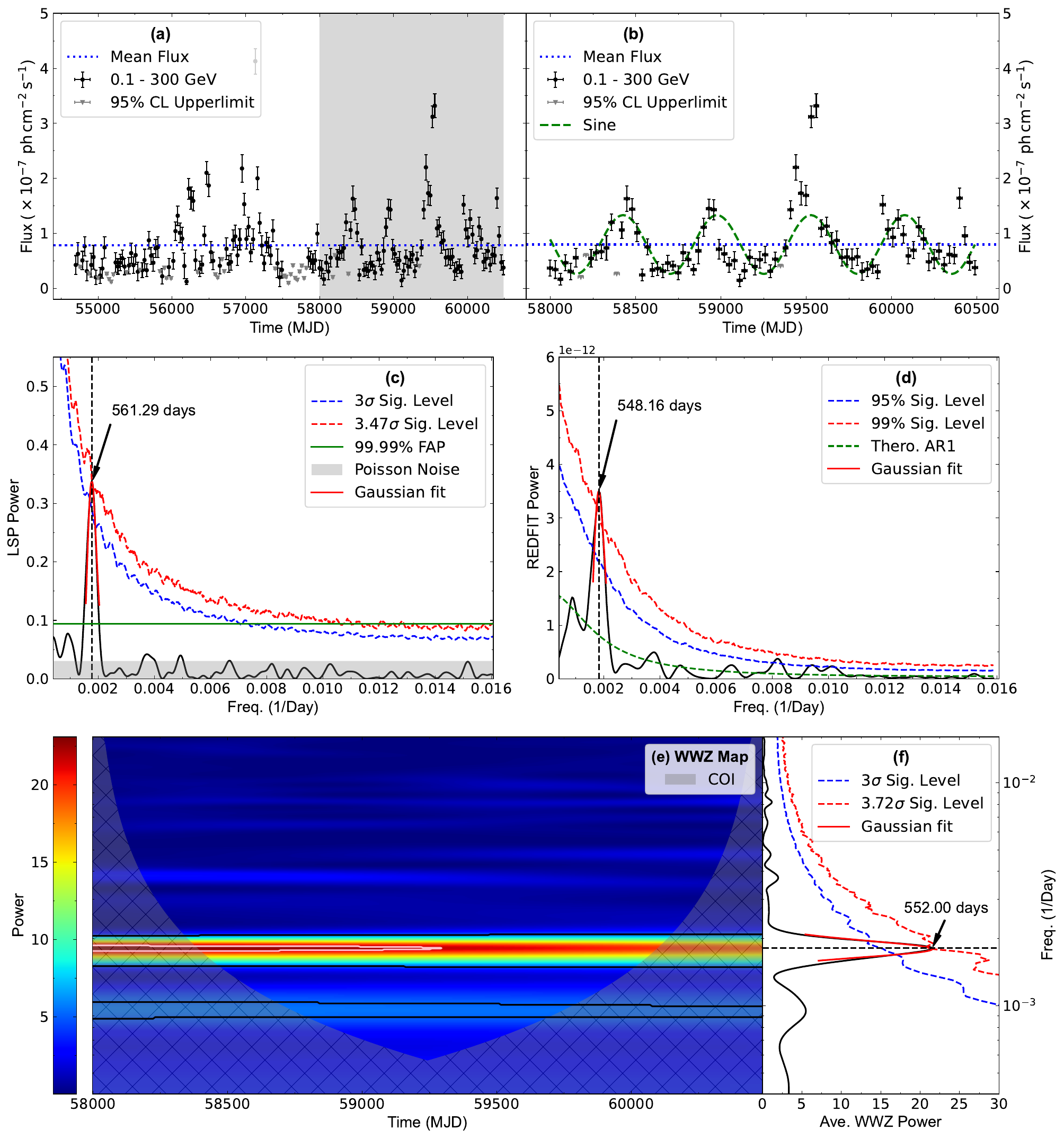}
\caption{
Panel (a): The \textit{Fermi}-LAT light curve of 4FGL J0309.9-6056 for $\sim$ 16 years (MJD 54683$-$60503). 
The gray inverted triangles represent the 95\% significance level upper limits.  
Panel (b): The shaded section (MJD 57983$-$60503) from the left panel is enlarged. 
The blue dotted line represents the mean flux, 
and the green solid line shows a sine function with a mean period of approximately 550 days derived from LSP, REDFIT, and WWZ methods.}
Panel (c): The LSP results for the period MJD 57983$-$60503 with peak value of 561.29 $\pm$ 74.15 days. 
The blue and red dotted lines represent the 3.00$\sigma$ and 3.47$\sigma$ local significance levels. 
The green solid line indicates the 99.99\% false-alarm probability, and the red solid line is the Gaussian function fitted to the peak.
Furthermore, the Poisson noise level is calculated to be $\sim$ 0.031, represented in a gray region.
Panel (d): Results of the periodicity analysis by the REDFIT program with a peak at 548.16 $\pm$ 83.04 days for the period MJD 57983$-$60503. 
The black line is the PSD calculated by REDFIT, the red dashed line represents the 99\% significance levels by estimating the red noise background, and the green dashed line is the theoretical AR1 spectrum.
The red solid line is the Gaussian function fitted to the peak.
Panel (e): The WWZ power spectrum map for the period MJD 57983$-$60503.
The gray region is the cone of influence (COI).
panel (f): The black solid line shows the time-averaged WWZ. 
The blue dashed curves represent the 3.00$\sigma$ local significance levels. The red dashed curve represents the 3.72$\sigma$, which passes the peak value of 552.00 $\pm$ 65.66 days. 
The red solid line is the Gaussian function fitted to the peak.
\label{fig:QPO_0.1_300}
\end{figure*}

The Weighted wavelet Z-transform \citep[WWZ;][]{Foster1996} 
can transform data into the time domain and frequency domain and convolute the light curve with the kernel related to time and frequency.
It can get the power intensity of periodic feature to search the periodicity by decomposing the signal into the frequency time space, and study its duration period.
The Morlet kernel is defined as: 
\begin{equation}
    f(\omega(t-\tau)) = e^{i \omega (t-\tau) - c \omega^2 (t-\tau)^2 } {\rm ,}
\end{equation} 
where $\omega$ is the angular frequency, $\tau$ is the time translation parameter, and $c$ is the window decay rate.
Then, the WWZ power is given by: 
\begin{equation}
W[\omega,\tau;x(t)] = \omega^{1/2}\int{x(t) f^* [\omega(t-\tau)]\mathrm{d}t}{\rm ,} 
\end{equation}
where the $f^*$ is the complex conjugate of the Morlet kernel $f$ and $x(t)$ is the light curve. 
More information concerning the WWZ method can be found in \citet{Foster1996}. 
We used a WWZ analysis python package\footnote{\url{https://github.com/skiehl/wwz}} to obtain the color map of the WWZ power spectrum and the average power in a function of frequency.
We set the frequency range of of $f_{\rm min}=\frac{1}{T}$ to $f_{\rm max} = \frac{N}{2 T}$ with a step of 0.00015 and used $c$= 0.001.
We also calculated the cone of influence (COI) to account for edge effects arising from the finite length of the data.
The COI marks the region of the wavelet power spectrum where edge effects become significant and the results are less reliable.
The results are shown in panels (e) and (f) of Figure \ref{fig:QPO_0.1_300}, and the time-average WWZ power gives the peak at 552.00 days.
The corresponding periodicity uncertainty is estimated from the FWHM of the Gaussian function fitted to the peak.

\subsection{Significance estimation}\label{sigma_estimation}

As mentioned above, we employed three methods to analyze QPO signals. 
However, the light curves of most AGNs exhibit frequency-dependent red-noise characteristics, where both random flares and sampling instability can lead to the appearance of red noise, potentially generating false QPO signals.
To quantify the significance of the observed periodicity detected by the LSP and WWZ methods, we applied the method developed by \citet{Emmanoulopoulos2013}, which builds upon the approach by \citet{Timmer1995}. 
This method relies on the same properties of power spectral density (PSD) and probability distribution function (PDF) for the original light curve. 
Subsequently, the Python code \texttt{DELCgen} \citep{Connolly2016ascl} was used to generate $10^5$ simulated light curves, which were then resampled to match the observational sampling, allowing us to assess the significance of the periodicity.
The statistical significance derived from this procedure represents the local significance, which quantifies the significance level of the peak of the detected period at this specific period. 
However, without prior knowledge of the location of the peaks, it is more robust to check for a “global significance”.
Given that the search spans a wide range of frequencies, the possibility of detecting a spurious peak increases, which is also known as the “look-elsewhere effect” or “multiple comparison problem” in statistics \citep{Bell2011MNRAS411}.
Therefore, the global significance estimates the significance of observing such a significant peak at any frequency within the search range, without prior knowledge of the peak location.

\begin{figure}
\centering
\includegraphics[width=3.4 in]{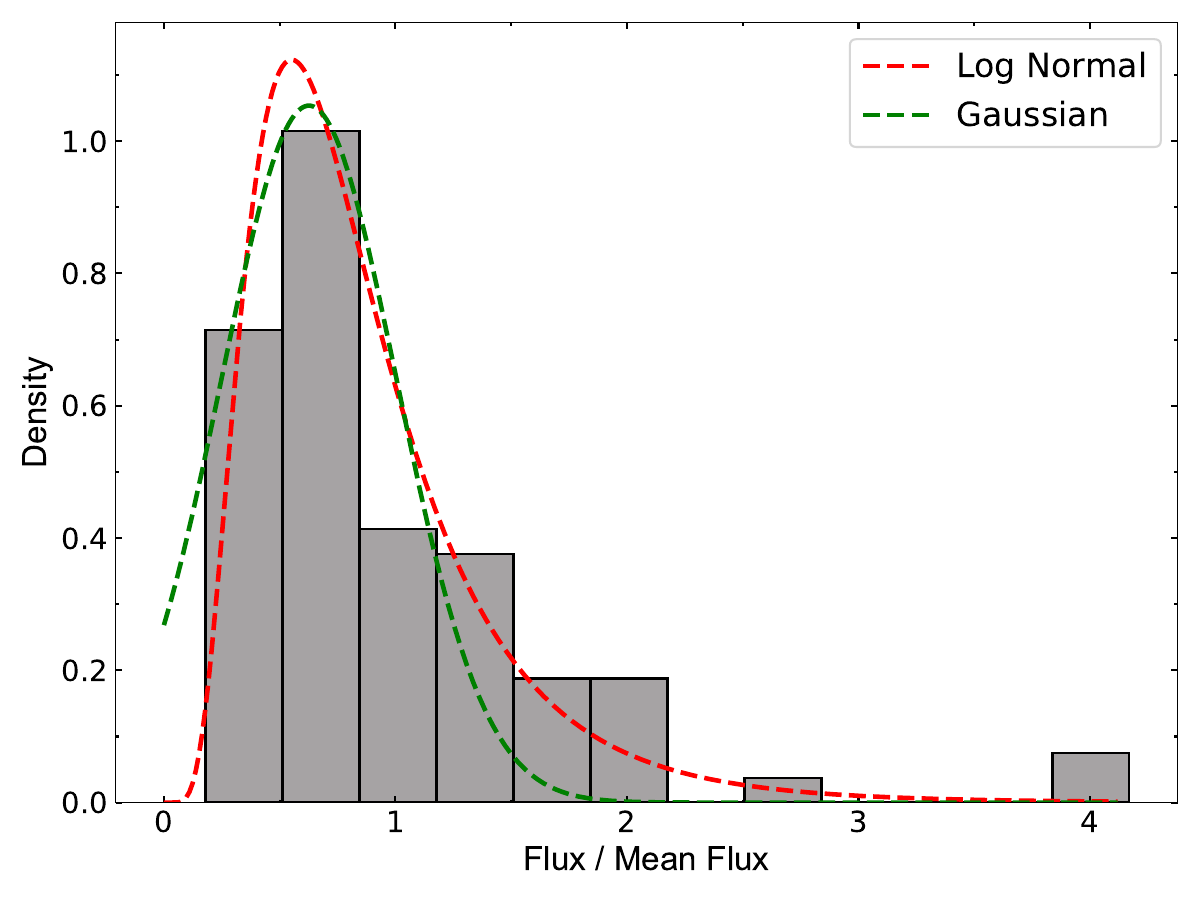}
\caption{
Histogram of $\gamma$-ray light curves for 4FGL J0309.9-6056. The red dashed line represents the log-normal fit; the green dashed one is the Gaussian fit.
}
\label{fig:PDF}
\end{figure}

We used the power-law model $P(f) \propto f^{-\beta} + P_{\rm noise}$ to effectively model the red-noise PSD of the original light curve \citep{Uttley2002}, where $\beta>0$ is the power-law spectral slope, and $P_{\rm noise}$ represents the Poisson noise contribution.
The Poisson noise is defined as
\begin{equation}
    P_{\rm noise}=\frac{2 T }{N^{2} \mu^{2}} \overline{F_{\rm err}^{2}},
\end{equation}
where $N$ is the total number of measurements, $\mu$ is the mean flux, $T$ is the total period of observation, and $\overline{F_{\rm err}^{2}}$ is the mean square of the flux uncertainties. 
We estimate the power-law spectral slope using the PSRESP method, which provides the ``success fraction" as a measurement of the goodness of fit \citep{Uttley2002, Chatterjee2008ApJ689, Max-Moerbeck2014MNRAS445}.
In this method, a total of $M = 1000$ artificial light curves with red-noise characteristics were generated for each trial power-law slope $\beta$, ranging from 0.5 to 2.5 in steps of 0.1, using the Monte Carlo approach with \citet{Timmer1995}. 
Each simulated light curve was then resampled to match the observational sampling and processed to compute its PSD in the same way as the observed data.
The PSRESP method evaluates how well the assumed PSD model reproduces the observed PSD by comparing the distribution of simulated PSDs with the observed one using a $\chi^2$-like function, which is defined as
\begin{equation}
\chi^2_{\rm {obs}} = \sum_{\nu = \nu_{\min}}^{\nu_{\max}} 
\frac{(\mathrm{PSD}_{\rm {obs}} - \overline{\mathrm{PSD}}_{\rm {sim}})^2}
{(\Delta \mathrm{PSD}_{\rm {sim}})^2}   ,
\end{equation}
and 
\begin{equation}
\chi^2_{{\rm dist}, i} = \sum_{\nu = \nu_{\min}}^{\nu_{\max}} 
\frac{(\mathrm{PSD}_{{\rm sim}, i} - \overline{\mathrm{PSD}}_{\rm {sim}})^2}
{(\Delta \mathrm{PSD}_{\rm {sim}})^2}, 
\end{equation}
where $\overline{\mathrm{PSD}}_{\rm {sim}}$ is the average of $\mathrm{PSD}_{{\rm sim}, i}$ and $\Delta \mathrm{PSD}_{\rm {sim}}$ is the standard deviation of $\mathrm{PSD}_{{\rm sim}, i}$.
The ``success fraction" is then determined by $\frac{m}{M}$ of the searching trial slopes, where $m$ is the count of the number of $\chi^2_{{\rm dist}, i}$ for which $\chi^2_{\rm {obs}}$ is smaller than $\chi^2_{{\rm dist}, i}$. 
The power-law slope distribution is shown in Figure \ref{fig:PDF}, giving a reliable estimate of the intrinsic spectral slope as $\beta_{\rm opt}$ = 1.26 $\pm$ 0.28. 
This value corresponds to the peak of a Gaussian function fitted to the distribution, with the associated uncertainty derived from the FWHM of the Gaussian.
The PDF was conducted from the flux distribution histogram, as shown in Figure \ref{fig:PDF}. 
The Shapiro–Wilk statistics were applied to assess whether the original light curve originated from a Gaussian or a log-normal distribution \citep{Shapro1965Biometrika}. 
The Shapiro–Wilk $p$-values are 1.86 $\times$ 10$^{-9}$ and 0.45 for linear-scale and log-scale distribution tests, respectively, indicating flux distribution follows the log-normal distribution. 
The presence of a log-normal flux distribution suggests that the variability is driven by a nonlinear multiplicative mechanism. 
In AGNs, such a distribution is often linked to fluctuations propagating through the accretion disk, where perturbations in the mass accretion rate multiply as they propagate inward, leading to a log-normal distribution of the flux \citep{Uttley2005MNRAS359}.
In the case of blazars, where $\gamma$-ray variability is primarily associated with non-thermal radiation from the jet, the log-normal distribution may reflect multiplicative perturbations originating in the accretion disk and subsequently propagating into the jet. 
Additionally, $\gamma$-ray variability in blazars could arise from variations in jet instabilities, magnetic fields, particle densities, or seed photon fields, all of which can contribute to the log-normal flux distribution \citep{Bhatta2020ApJ891}.

Finally, the local significance of the periodicity was estimated from the percentile distributions of the LSP and WWZ power at each frequency, derived from the $10^5$ simulated light curves.
The blue and red dashed curves in Figure \ref{fig:QPO_0.1_300} represent the local significance levels of the LSP and WWZ methods.
Furthermore, we estimated the global significance of the LSP and WWZ peaks using the approach described in \citet{O'Neill2022ApJ926L}.
As a result, we identified a periodic signal of 561.29 $\pm$ 74.15 days with a local significance of 3.47$\sigma$ and a global significance of 2.30 $\sigma$ using the LSP method, a signal of 548.15 $\pm$ 83.04 days exceeding a 99\% significance level using REDFIT, and a signal of 552.00 $\pm$ 65.66 days with a local significance of 3.72$\sigma$ and a global significance of 2.72 $\sigma$ based on the average WWZ power. 
All three methods consistently detected a QPO with a period of approximately 550 days in the $\gamma$-ray band, which corresponds to the mean value derived from these methods. 

\begin{figure}
\centering
\includegraphics[width=3.4 in]{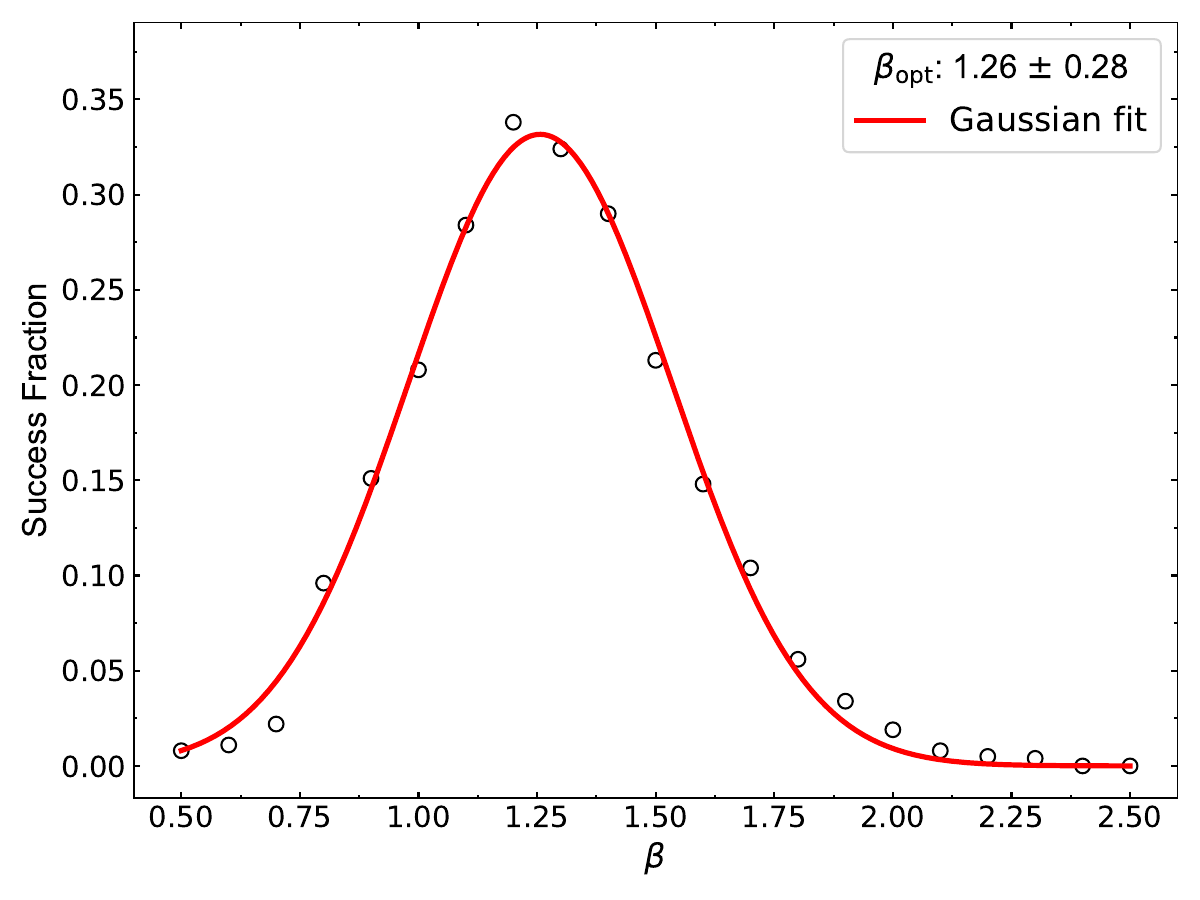}
\caption{
Power-law slope distribution as obtained using the PSRESP method.
The best-fitting value of $\beta_{\rm opt}$ is obtained from the peak of a Gaussian function fit, and its associated uncertainty is taken from the FWHM of the Gaussian. 
}
\label{fig:PSD}
\end{figure}

\begin{table}
 \caption{Fitting parameters for flux distributions of Figure \ref{fig:PDF}. RSS is the residual sum of squares.}
 \label{tab:pdf_fit_results}
 \begin{tabular}{lccc}
  \hline
    & $\mu$ & $\sigma$ & RSS \\
  \hline
  Gaussian & 0.63 & 0.38 & 1.62 \\
  Log-normal & -0.29 & 0.55 & 1.18 \\ 
  \hline
 \end{tabular}
\end{table}

\subsection{Gaussian Process modeling}
The three methods used above, LSP, WWZ, and REDFIT, all analyze periodicity in the frequency domain. 
Gaussian Process (GP) modeling treats the observed variability as a realization of a stochastic process and allows for flexible modeling of correlated noise and intrinsic variations, which can be used to analyze periodicity in the time domain.

To model the light curve, we construct a GP composed of a sum of two stochastically-driven damped harmonic oscillator (SHO) terms, each capturing variability on different timescales.
Each SHO term is described by the stochastic differential equation:
\begin{equation}
\left[ \frac{d^2}{dt^2} + \frac{\omega_0}{Q} \frac{d}{dt} + \omega_0^2 \right] y(t) = \epsilon(t),
\end{equation}
where $\omega_0$ is the undamped natural frequency, $Q$ is the quality factor (which controls the sharpness of the resonance), and $\epsilon(t)$ is a white noise process. 
The corresponding power spectral density is given as:
\begin{equation}
S(\omega) = \sqrt{\frac{2}{\pi}} \frac{S_0 \omega_0^4}{\left( \omega^2 - \omega_0^2 \right)^2 + \omega^2 \omega_0^2 / Q^2},
\end{equation}
where $S_0$ is the power normalization. 

\begin{figure*}
\centering
\includegraphics[width=5.3 in]{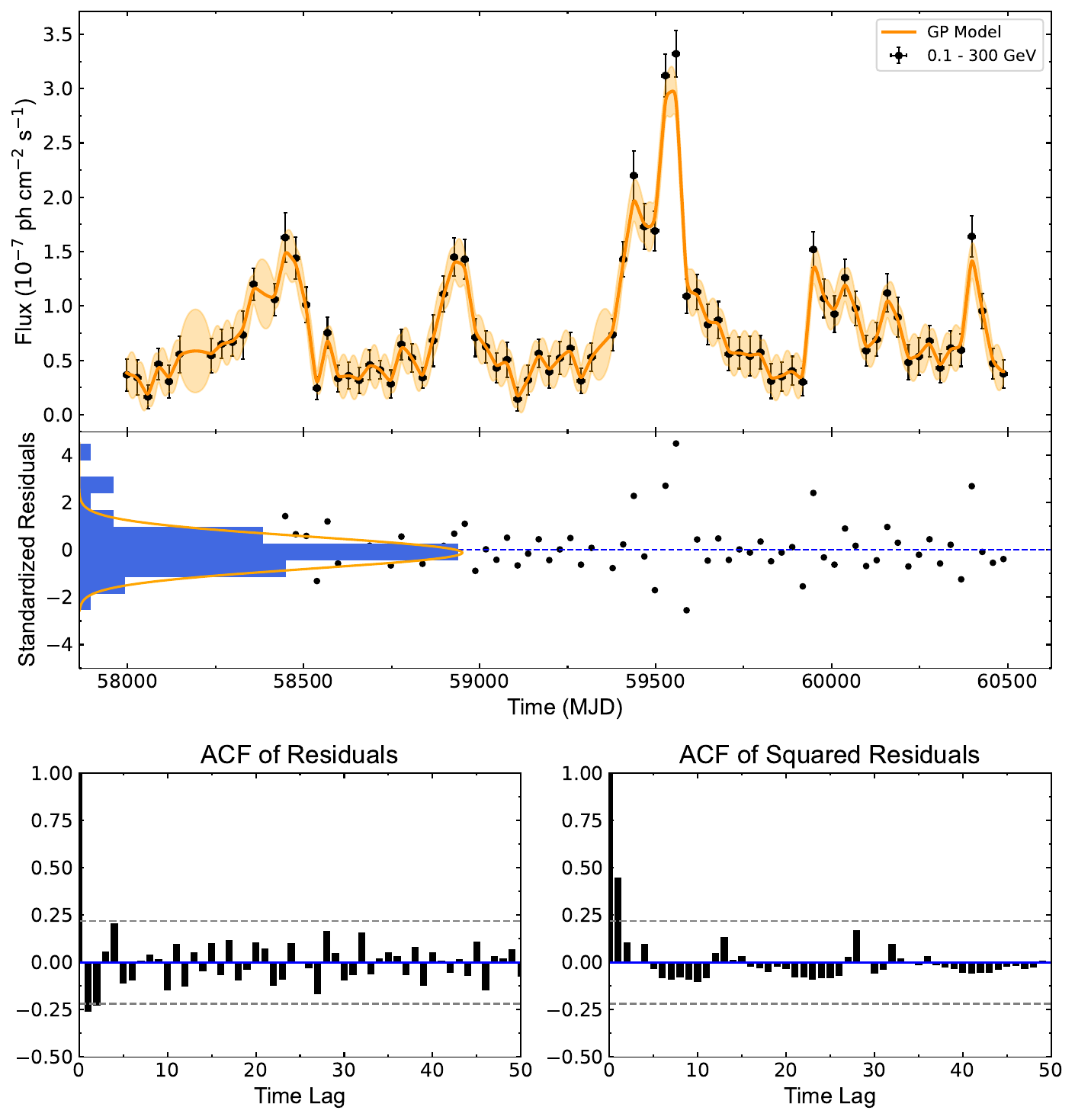}
\caption{
% Results of the Gaussian Process. 
Gaussian Process modeling of the 0.1–300 GeV $\gamma$-ray light curve for the period MJD 57983$-$60503, obtained using the SHO$\times$2 model.
Top panel: Observed data and the best-fit profile including the 1$\sigma$ confidence interval. 
Middle panel: Standardized residuals of the Gaussian Process fit as a function of time.
Bottom panels: Autocorrelation function (ACF) of the residuals (left) and squared residuals (right). 
}
\label{fig:Gaussian_Process}
\end{figure*}

\begin{figure*}
\centering
\includegraphics[width=3.5 in]{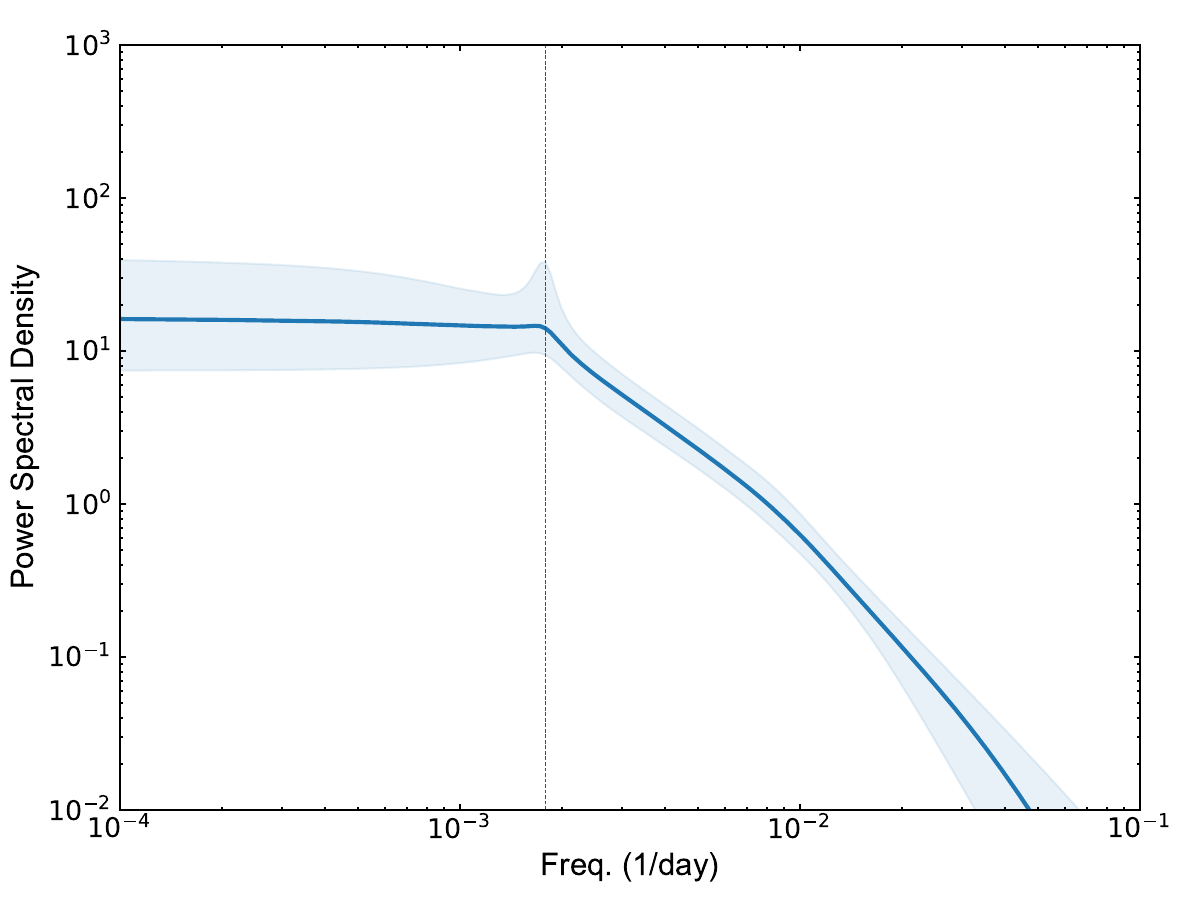}
\caption{
Power spectral density of the Gaussian Process modeling obtained using the SHO$\times$2 model. 
The shaded regions correspond to 1$\sigma$ confidence interval. 
}
\label{fig:model_PSD}
\end{figure*}

We use the \texttt{celerite}\footnote{\url{https://celerite.readthedocs.io/en/stable/}} package \citep{Foreman-Mackey2017AJ154} to construct the light curve variability, in which the parameters are expressed in the natural logarithmic space.
Parameter estimation is carried out using Markov Chain Monte Carlo (MCMC) with \texttt{emcee} sampler, generating 32$\times$20000 samples and discarding the initial 32$\times$2000 as burn-in.
The remaining samples are then used to derive the posterior estimates.
The modeled light curve and the fitted values of model parameters are presented in Figure \ref{fig:Gaussian_Process} and Table \ref{tab:GP_results}, respectively. The posterior distributions for the model parameters are shown in Figure \ref{fig:Cornor_GP} in the Appendix. Additionally, the power spectral density (PSD) of the model is shown in Figure \ref{fig:model_PSD}.
And it shows a peak at the frequency of $\sim$0.00178 {day$^{-1}$} (561.79 days).

Residual analysis using the Shapiro–Wilk test confirms that the residuals are consistent with normality ($p$=0.26), suggesting the model effectively accounts for the intrinsic variability.
Moreover, the autocorrelation function (ACF) and squared ACF of the residuals remain within the 95\% confidence interval, suggesting that the model successfully captures the temporal correlation structure in the data.

As a result, the GP modeling also revealed a period consistent with the QPO period of $\sim$550 days identified by the Fourier-based methods.
The detailed results obtained from each method are summarized in Table \ref{tab:periodicity_results}.

\begin{table}
 \caption{Parameters of Gaussian Process modeling}
 \label{tab:GP_results}
 \begin{tabular}{lccc}
  \hline
  Model & $\ln S_{0}$ & $\ln Q$ & $\ln \omega_{0}$ \\
  \hline
    \multirow{2}{*}{SHO $\times$ 2} & 2.83$^{+0.95}_{-1.31}$ & -1.37$^{+0.80}_{-0.79}$ & --2.85$^{+0.52}_{-0.40}$  \\
                  & -0.35$^{+2.50}_{-2.84}$ & 0.88$^{+2.44}_{-2.89}$ & -4.49$^{+0.23}_{-0.23}$ \\
  \hline
  Prior&
  (-5, 5) &
  (-5, 5) &
  (-5,-2) \\
  \hline
 \end{tabular}
\end{table}

\begin{table*}
 \caption{Detected periods from different methods.}
 \label{tab:periodicity_results}
 \begin{tabular}{llcc}
  \hline
  Method & Period & Local significance & Global significance \\
  \hline
  Lomb-Scargle Periodogram (LSP) & 561.29 $\pm$ 74.15 & 3.47$\sigma$ & 2.30$\sigma$ \\
    Weighted Wavelet Z-transform (WWZ) & 552.00 $\pm$ 65.66 & 3.72$\sigma$ & 2.72$\sigma$ \\
    REDFIT & 548.16 $\pm$ 83.04  & -- & -- \\
    % Seasonal ARIMA (SARIMA) & 510.0 $\pm$ 15.0 \\
    Gaussian Process & 561.79 $^{+130.54}_{-129.20}$  & -- & --  \\
  \hline
 \end{tabular}
\end{table*}

\section{Discussion} \label{sec: dis}
Building upon the findings in the previous section, both Fourier-based methods and GP modeling consistently identified a QPO with a characteristic timescale of approximately 550 days.
\subsection{Extended analysis throughout the entire duration}
\begin{figure*}
    \centering
    \includegraphics[width=6.7 in]{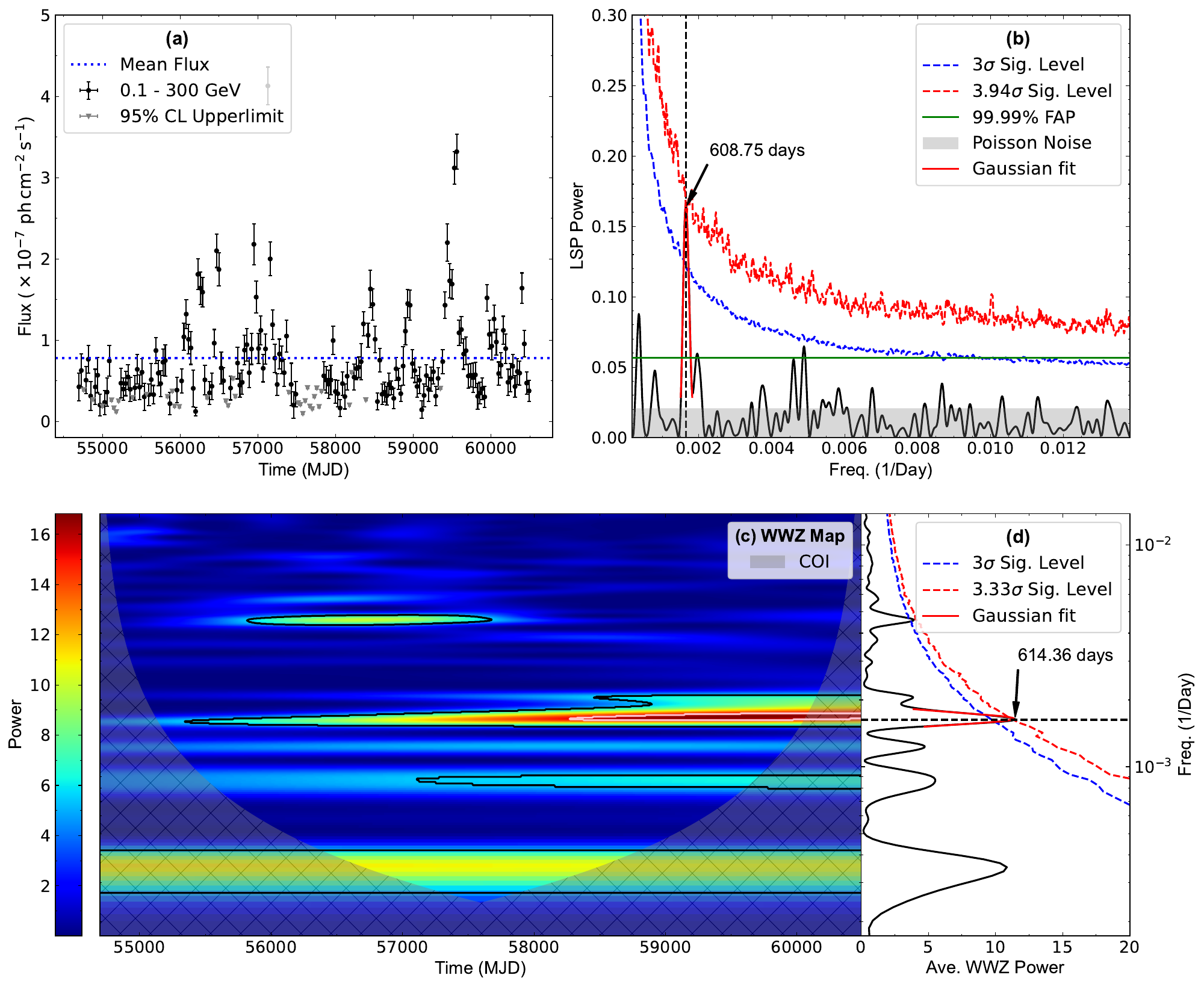}
    \caption{
    Panel (a): $\gamma$-ray LC for the entire time of \textit{Fermi}-LAT observations (MJD 54683$-$60503). 
    The blue dotted line is the mean flux.
    Panel (b): The LSP results for MJD 54683$-$60503. There is a peak at 608.75 $\pm$ 42.60 days with more than 3$\sigma$ local significance. 
    The other labels are the same as Figure \ref{fig:QPO_0.1_300}.
    Panels (c) and (d) show the WWZ power spectrum map and the time-averaged WWZ power for the period MJD 54683$-$60503. A significant peak is observed at 614.36 $\pm$ 54.74 days, exceeding the 3$\sigma$ local significance level.
    }
    \label{fig:QPO_all}
\end{figure*}

We extended our search to cover the entire time span of \textit{Fermi}-LAT observations and applied the LSP method to analyze the periodicity over MJD 54683–60503. 
As shown in Figure \ref{fig:QPO_all}, a periodic signal with a timescale of approximately 600 days is detected, with a local significance exceeding 3$\sigma$ and a global significance of 2.49$\sigma$.
We also perform the WWZ analysis for the entire duration to examine the time localization of approximately 600-day QPO with a significance exceeding 3$\sigma$ and a global significance of 2.24$\sigma$. 
The result confirms that this periodic signal exists nearly in the full period, as shown in panels (c) and (d) of Figure \ref{fig:QPO_all}. 
These significances were estimated following the approach as described in Section \ref{sigma_estimation}.  
This result is consistent with previous findings, but now confirmed over a longer observational baseline.

\subsection{Time lag between optical and gamma-ray band}

\begin{figure*}
    \centering
    \includegraphics[width=6.8 in]{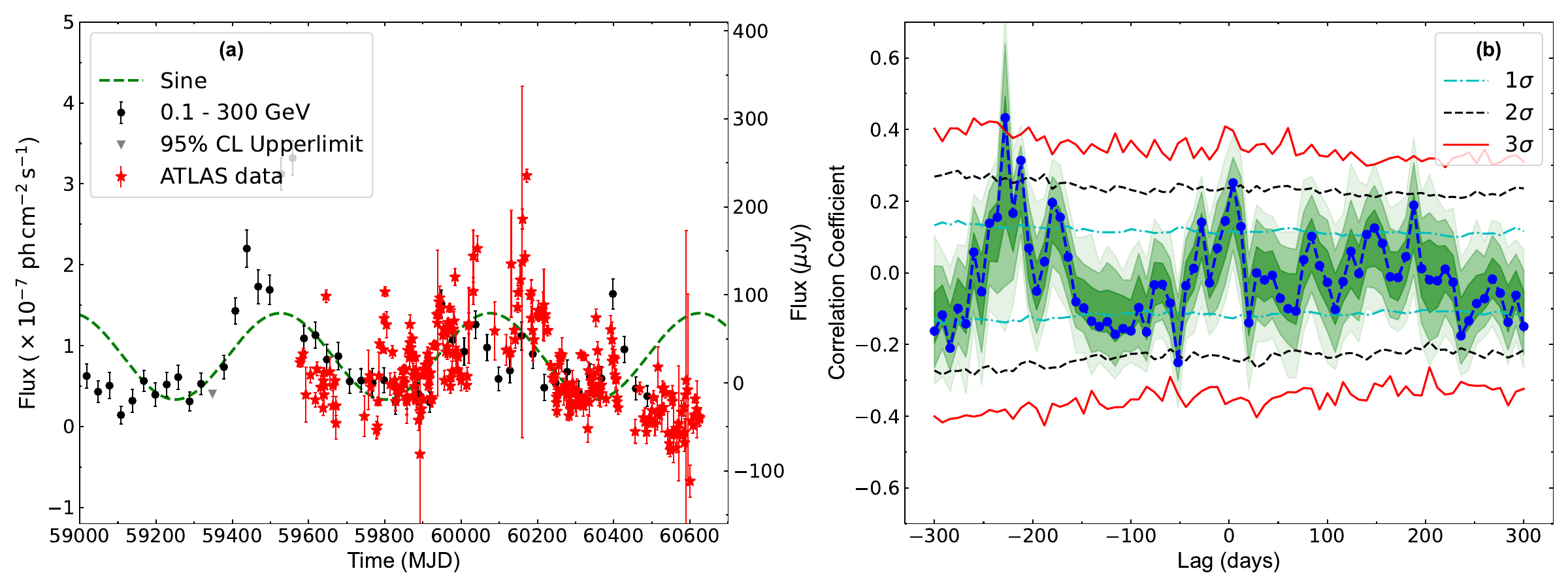}
    \caption{
    Panel(a): The red point represents the light curve of this source in the optical band. The black dots represent the light curve of the source quasi-simultaneously with the optical band.
    The optical light curve was binned at 1-day intervals. 
    For the $\gamma$-ray data, a 30-day binning was used for visualization here, while a 5-day binning was adopted specifically for the DCF analysis.
    The green line is the sine curve derived from panel (b) of Figure \ref{fig:QPO_0.1_300}.
    Panel(b): The result of cross-correlation between $\gamma$-ray and optical bands with a bin size of 8 days; similar results are obtained for other bin sizes.
    The blue, black, and red lines represent the 1.0, 2.0, and 3.0$\sigma$ significance levels.
    The peak at -228 days indicated that the optical band is ahead of the $\gamma$-ray band.
    }
    \label{fig:QPO_optical_timelag}
\end{figure*}

We also extended our search to other wavelengths and found that the Asteroid Terrestrial-impact Last Alert System (ATLAS) observation contributed to this investigation. 
After data collection, automated image processing was performed, including photometric and astrometric calibration using the RefCat2 reference catalog \citep{Tonry2018PASP130, Tonry2018ApJ867}. 
A reference image was then subtracted to identify transient events. 
Detected sources in the difference images were filtered through a transient discovery pipeline \citep[the ATLAS Transient Server;][]{Smith2020PASP132}. 
For this study, we queried the ATLAS forced photometry service for data spanning from MJD 59577 to 60625. 
We plotted the ATLAS data together with the $\gamma$-ray data and the sine function in Panel (a) of Figure \ref{fig:QPO_optical_timelag}.
We noticed that the optical ATLAS data appear to coincide with the predicted $\gamma$-ray sine function in the range of MJD 59800-60400, and this coincidence suggests that the QPO signal may also appear in the optical band, although we lack sufficient optical data to perform a detailed periodic analysis.
Given the similarity in flux variation between the $\gamma$-ray and optical bands in Panel (a) of Figure \ref{fig:QPO_optical_timelag}, we analyzed the cross-correlation between the $\gamma$-ray and optical flux using discrete correlation function \citep[DCF;][]{Edelson1988APJ333} with the \texttt{MUTIS}\footnote{\url{https://mutis.readthedocs.io/en/latest/}} package.
During the DCF analysis, the optical data were binned into 1-day intervals, and the $\gamma$-ray data were binned into 5-day intervals.
Note that a shorter time interval of the $\gamma$-ray light curve can reveal more detailed DCF structures, but this comes with increased flux errors and reduced TS values, which can decrease the quality of the DCF result.
A compromised 5-day interval $\gamma$-ray light curve data was chosen to avoid numerous upper limit data points, which could significantly reduce the quality of the DCF result and obtain a trustable DCF result.
The statistical significances and the uncertainties of the DCF correlation were estimated using a Monte Carlo approach by generating $N = 2000$ synthetic light curves for each signal.
The generation process used the Lomb-Scargle to compute the PSD and the non-uniform Fourier transform to reconstruct the signals with similar PSD, mean, and standard deviation. 
Detailed information can be found in the description of \texttt{MUTIS}.
The 8-day DCF bin size is used to calculate the DCF correlation, which is presented in Panel (b) of Figure \ref{fig:QPO_optical_timelag}, showing a time-lag of -228 days with 3.5$\sigma$.
We also tested the different DCF bin sizes of 10, 12, 15, and 20, finding a consistent $\sim$220-day time lag with significance levels of 1.9$\sigma$, 2.5$\sigma$, 3.2$\sigma$, and 3.2$\sigma$, respectively.
The time lag between optical and $\gamma$-ray band is very likely to exist, and this lag provides valuable constraints for QPO models.

In the model of jet helical structure, the blob helically moving forward along the jet could cause a periodically changing viewing angle, Doppler factor, and further flux variation.
On one hand, this model usually yields a QPO time scale range from a few days to months, and these blobs are likely generated during flares and last for only a short time scale \citep{Rieger2004ApJ, Rani2009ApJ, Zhou2018NatCo, Banerjee2023MNRAS, Chen2024MNRAS}.
On the other hand, we found an optical-$\gamma$-ray time lag suggesting separated emission regions for these two bands, and in conflict with the single moving blob in the helical jet model.
Unless we assume a complicated helical model of two separated blobs simultaneously moving in the jet.
Thus, this model is less promising for interpreting the QPO signal in this work.
In the following, we focus on the binary supermassive black hole (SMBH) system and the precession of the relativistic jet.

\subsection{Binary supermassive black hole system }
The binary supermassive black hole system assumes that the center of the galaxy consists of two supermassive black holes, providing an important framework for explaining the QPO phenomenon.
There are two different explanations:
\textit{lighthouse model} \citep{Villata1998MNRAS,Qian2007ChJAA} and \textit{accretion model} \citep{Lehto1996ApJ,Valtonen2006ApJ,Qian2007ChJAA,Fan2014ApJS}.
In the framework of \textit{lighthouse model}, \citet{Villata1998MNRAS} suggested that both black holes in the binary system generate relativistic jets that are bent significantly in different directions.
In the course of the binary’s orbit motion, the directions of the bent parts of the jets from the two black holes rotate with the orbital period, resulting in periodic double-peak flares.
If so, the $\gamma$-ray light curve would exhibit a distinct double-peak feature.
However, we do not observe a clear double-peak feature (see panel (a) of Figure \ref{fig:QPO_0.1_300}).
Thus, this model is less likely to explain the QPO signal in this work.

\textit{Accretion model} is that the secondary black hole crossing the accretion disk of the primary black hole can increase the accretion rate.
For the binary supermassive black system, the total mass of the binary is $M_{\rm tot}=M_{\rm p}+M_{\rm s}$, where $M_{\rm p}$ is the mass of the primary BH and $M_{\rm s}$ is the mass of the secondary BH.
The orbital period of the binary $P$ can be calculated by Kepler's law 
\begin{equation}
    (a_1+a_2)^{3}=\frac{GM_{\rm tot}}{4\pi^{2}} P^{2} {\rm , }
\end{equation}
% \b{where $r_{ps}$ is the sum of the semi-major axes}
where $ a_1$ and $ a_2$ are semi-major axes.
It can be equivalent to the following \citep[e.g.,][]{Fan2014ApJS,Fan2021ApJS..253...10F}
\begin{equation}
    P \sim 1.72 M_{8}^{-1/2} r_{16}^{3/2} \left(1+\frac{M_{\rm s}}{M_{\rm p}} \right)^{-1/2} \, {\rm yr. }
\end{equation}
The $M_8$ is the primary black hole mass in units of ${10{^8}}{M_{\odot}}$ and
% \b{$r_{16}$ is $r_{ps}$ in units of $10^{16}$ cm}.
$r_{16}= a_1+a_2$ is in units of $10^{16}$ cm.
And the $P$ can be calculated by observed period $P_{\rm obs}$ as
\begin{equation}
    P =\frac{P_{\rm obs}}{1+z}  {\rm . }
\end{equation}
Taking $M_{\rm p} = 10^{8.87} M_{\odot}$ \citep{Shaw2012ApJ}  
% from Mg \uppercase\expandafter{\romannumeral2} broad emission line 
% \citep{Shaw2012ApJ}
% , that is $M_8=10^{0.87 \pm 0.27} M_{\odot}$.
, $\frac{M_{\rm s}}{M_{\rm p}} \sim 0.001$ and the observed period 550 days, 
% the $P_{\rm ps}$ can be calculated as 221.8 day in the rest frame and
we obtain $r_{16} = 1.044$ (3.383 milli-parsec).
The orbiting and merger of the binary supermassive black hole would generate a stochastic nHz GW background \citep{Sesana2013MNRAS}.
For the case of our source, we expect the gravitational waves at $f=2/P_{\rm obs} \sim $ 42.1 nHz.
For a quasi-circular orbit, the gravitational waves strain is given by \citep{maggiore_gravitational_2007}
\begin{equation}
h=\frac{2(GM_\oplus)^{5/3}(\pi f)^{2/3}}{c^4D} {\rm , }
\end{equation}
where the luminosity distance of this source is 11.00 Gpc and the observed-frame chirp mass is 
\begin{equation}
M_\oplus=\frac{(1+z)(M_sM_p)^{3/5}}{(M_s+M_p)^{1/5}} {\rm . }
\end{equation}
We can get the gravitational waves strain $h=1.8 \times 10^{-19}$. 
Current gravitational wave detectors LIGO \citep[10 Hz$-$10 kHz;][]{LIGO2015CQGra..32g4001L} and Virgo \citep[10 Hz$-$up to a few kHz;][]{Accadia2012JInst} cannot detect gravitational waves in this frequency.
Currently, the Pulsar Timing Array (PTA) is the only known effective method to detect GWs in the nHz band.
There are several major PTAs: Parkes Pulsar Timing Array \citep[PPTA;][]{Manchester2008AIPC}, North American Nanohertz Observatory for Gravitational Waves \citep[NANOGrav;][]{Jenet2009arXiv0909}, Chinese Pulsar Timing Array \citep[CPTA;][]{Lee2016ASPC..502...19L,Xu2023RAA}, South Africa Pulsar Timing Array \citep[SAPTA;][]{Spiewak2022PASA...39...27S}, European Pulsar Timing Array \citep[EPTA;][]{Chalumeau2022MNRAS} and International Pulsar Timing Array \citep[IPTA;][]{Manchester2013CQGra..30v4010M}.
Specifically, based on the 15-year dataset of NANOGrav \citep{Agazie2023ApJ951L9A}, the sensitivity of NANOGrav \citep[$h \sim 10^{-14}$;][]{Agazie2023ApJ951L10A} is not enough for the gravitational waves strain of this source. 

We propose that it may be possible to distinguish binary black hole systems on a geometric scale.
The broad-line region (BLR) luminosity ($L_{\rm BLR}$) can be calculated using the following equation:
\begin{equation}
L_{\rm BLR}=L_{\rm line}\frac{<L_{\rm BLR}>}{L_{\rm line,frac}} {\rm , }
\end{equation}
where $L_{\rm line}$ denotes the emission-line luminosity and $L_{\rm line,frac}$ represents the luminosity ratio.
The luminosity ratios utilized are 77, 22, 34, and 63 for $\rm H{\alpha}$, $\rm H{\beta}$, $\rm Mg \, {\uppercase\expandafter{\romannumeral2}}$, and $\rm C \, {\uppercase\expandafter{\romannumeral4}}$ \citep{Celotti1997}.
Using the data from \citet{Shaw2012ApJ}, we calculated the BLR luminosity $\log L_{\rm BLR}$=44.88 erg $\rm s^{-1}$.
Assuming BLR covering factor is $0.1$, we got the accretion disk luminosity $\log L_{\rm disk}$ as 45.88 erg $\rm s^{-1}$.
The size of the BLR ($R_{\rm BLR}$), calculated by the equation $R_{\rm BLR}=10^{17}L^{1/2}_{\rm disk,45}$= $2.75 \times 10^{17}$ cm \citep{Ghisellini2008MNRAS387,Zhang2024ApJs27127Z}.
Compared to $r_{16}$, the size of the BLR is larger than the separation between the binary black holes, making it challenging to distinguish the binary black holes using the optical spectrum.
Consequently, confirming the binary black hole systems through optical observations remains extremely challenging.

\subsection{Jet precession}

The jet precession model emerges as the most promising explanation.
The precessing jet generates QPO signals in both the optical and $\gamma$-ray bands and the observed time lag between these bands reveals the distance between the optical and $\gamma$-ray emission regions. 
Jet precession can be induced by mechanisms such as a binary black hole system \citep{Katz1997APJ} or Lense-Thirring (LT) precession \citep{Lense1918PhyZ}.

\begin{figure*}
    \centering
    \includegraphics[width=5.0 in]{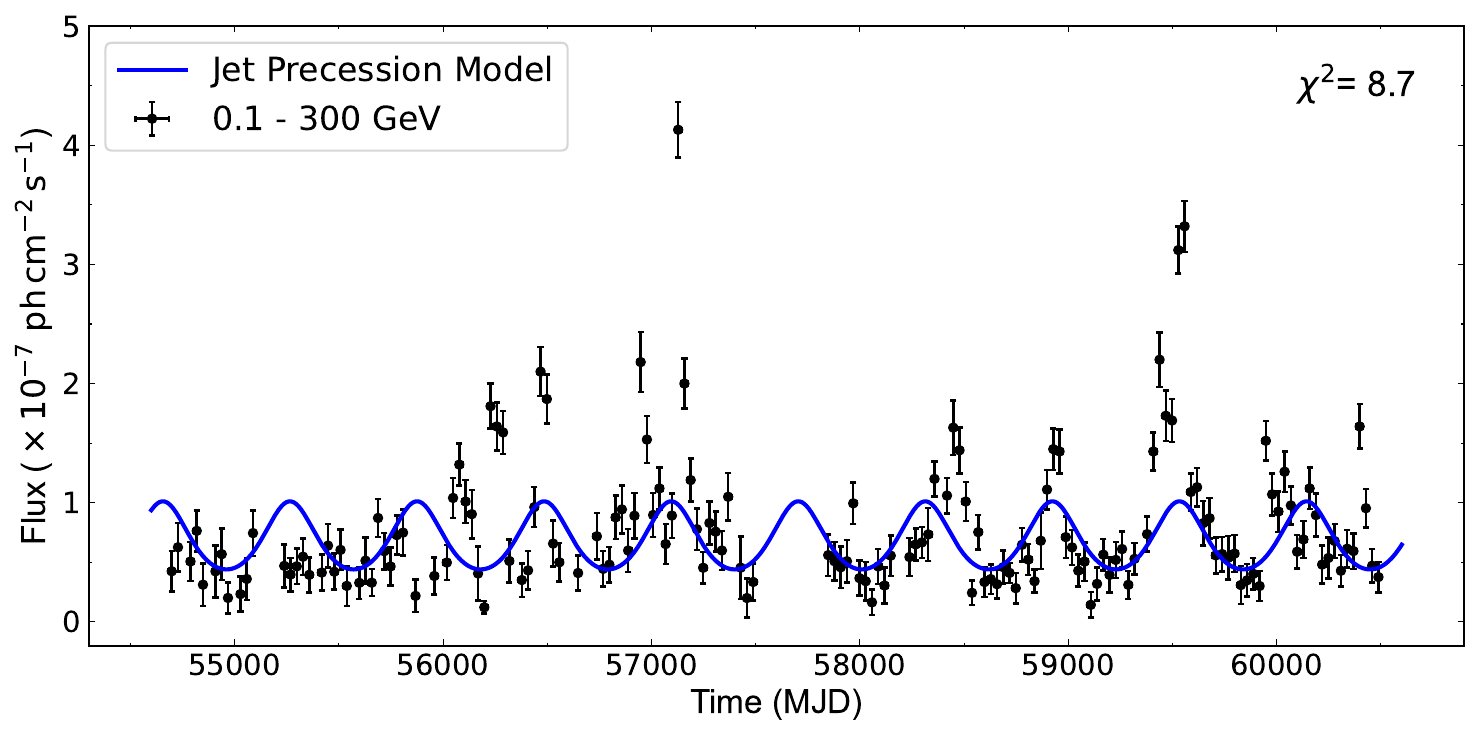}
    \caption{
    % This is the fitting jet precession model. 
    The fitting jet precession model. 
    The solid line represents the results of the precessing jet model. 
    }
    \label{fig:jet_precession}
\end{figure*}

Considering the jet precession model,
relativistic jet precessing goes around an axis and forms a conical surface with a precession angle $\Omega$.  
The cone axis forms an angle $\Phi_{0}$ with the direction of the line of sight and has a projected angle $\eta_{0}$ on the plane of sky \citep[e.g.,][]{Britzen2018MNRAS}. 
The time-dependent viewing angel ($\theta$) and the position angle ($\eta$) can be expressed by 
\begin{equation}
    \eta(t) = \arctan\frac{y}{x}    
    \label{Equ: eta}
\end{equation}
\begin{equation}
     \theta(t) = \arcsin \left( \sqrt{x^{2} + y^{2}} \right)  {\rm ,}
\end{equation}
with 
\begin{equation}
    x = A(t) \cos \eta_{0} - B(t) \sin \eta_{0}   \, ,
    y = A(t) \sin \eta_{0} + B(t) \cos \eta_{0}  
\end{equation}
and
\begin{equation}
    A(t) = \cos \Omega \sin \Phi_{0} + \sin \Omega \sin \omega (t - T_{0}) \cos \Phi_{0} \, ,
    B(t) = \sin \Omega \cos \omega (t - T_{0}) 
\end{equation}
where $\omega = 2 \pi / P_{\rm obs}$ is the angular velocity.  
The changing Doppler factor is obtained by $\delta = \Gamma \left( 1 - \beta \cos \theta (t) \right)^{-1} $, where
% the $\Gamma$ of the moving blob is 
$\Gamma = (1-\beta^2)^{-\frac{1}{2}}$ is the bulk Lorentz factor and $\beta = v_{\rm jet}/c$ is the bulk velocity.
Substituting above equations into $F=\delta^{3} F^{\prime}$,
we can obtain the varying observed flux due to the jet precession,
\begin{equation}
    F(t) = \frac{F^{\prime}}{\Gamma^{3}  \left[ 1- \beta  \cos \left( \arcsin \sqrt{x^{2} + y^{2}} \right)   \right]^{3} }   {\rm . }
    \label{Equ: fif}
\end{equation}
The term $\eta_{0}$ is completely canceled out in Equation (\ref{Equ: fif}), indicating that $\eta_{0}$ does not affect our result.
We modeled the observed light curve with the period of $P_{\rm obs}$ = 609.7 days, as shown in Figure \ref{fig:jet_precession}.

The light curve is well fitted by the jet precession model with the $\chi^{2}$ value between the model and the data as 8.7. 
The best-fit parameters are determined as follows: $\Omega$ = 2.0$^{\circ} \pm$0.6$^{\circ}$, $\Phi_{0}$ = 5.3$^{\circ} \pm$1.2$^{\circ}$, $\Gamma$ = 5.1$\pm$0.6, $F^{\prime}$=(11.6 $\pm$ 3.0) $\times 10^{-11}$ ph cm$^{-2}$ s$^{-1}$, $T_{0}$ = 55417.9 $\pm$ 6.1.
Our fitting results are reasonable, as $\Phi_0$ typically fluctuates by a few degrees, and previous studies have found the mean Doppler factor for FSRQs to be around 10 \citep{Hovatta2009, Fan2009PASJ, Liodakis2018}.

\section{Conclusion}
In this work, we report the detection of a QPO in the $\gamma$-ray band (0.1–300 GeV) of 4FGL J0309.9–6058 using $\sim$16 years of \textit{Fermi}-LAT observations. 
Through applying three Fourier-based time-series analysis methods, LSP, REDDIT, and WWZ, we consistently identified a QPO signal with a mean period of approximately 550 days.
Specifically, the LSP yielded a period of 561.29 $\pm$ 74.15 days with a local significance of 3.47$\sigma$ and a global significance of 2.30 $\sigma$; REDFIT indicated a period of 548.15 $\pm$ 83.04 days with a significance exceeding 99\%; and the WWZ method revealed a period of 552.00 $\pm$ 65.66 days with a local significance of 3.72$\sigma$ and a global significance of 2.72 $\sigma$.
In addition, Gaussian Process modeling independently produced a best-fit period of 560.66 days, consistent with the results obtained from Fourier-based methods.
We further extended our analysis to the full duration of the \textit{Fermi} observations, and the results consistently support the QPO signal, strengthening the reliability of our detection.
Additionally, we extended the QPO investigation to the optical band and found similar QPO behavior using ATLAS data.
However, more optical observations are necessary to firmly establish the QPO signal in that band. 
In addition, we detected a time lag of 228 days between the optical and $\gamma$-ray bands, suggesting the separated emission regions for optical and $\gamma$-ray emissions.
Considering the year-like timescale of the detected QPO and the time lag, we suggest that jet precession is the most plausible physical mechanism responsible for the QPO behavior for 4FGL J0309.9–6058.

\section*{acknowledgements}
H.B.X acknowledges the support from the National Natural Science Foundation of China (NSFC 12203034), the Shanghai Science and Technology Fund (22YF1431500), the science research grants from the China Manned Space Project (CMS-CSST-2025-A07), and the Shanghai Municipal Education Commission regarding artificial intelligence empowered research.
S.H.Z acknowledges support from the National Natural Science Foundation of China (Grant No. 12173026), the National Key Research and Development Program of China (Grant No. 2022YFC2807303), the Shanghai Science and Technology Fund (Grant No. 23010503900), the Program for Professor of Special Appointment (Eastern Scholar) at Shanghai Institutions of Higher Learning and the Shuguang Program (23SG39) of the Shanghai Education Development Foundation and Shanghai Municipal Education Commission.
J.H.F acknowledges the support from the NSFC U2031201, NSFC 11733001, NSFC 12433004, the Scientific and Technological Cooperation Projects (2020–2023) between the People’s Republic of China and the Republic of Bulgaria, the science research grants from the China Manned Space Project with No. CMS-CSST-2021-A06, and the support for Astrophysics Key Subjects of Guangdong Province and Guangzhou City.
This research was partially supported by the Bulgarian National Science Fund of the Ministry of Education and Science under grants KP-06-H38/4 (2019), KP-06-KITAJ/2 (2020) and KP-06-H68/4 (2022).

\section*{DATA AVAILABILITY}
The data presented in this article will be shared on reasonable request to the corresponding author.

\section*{Appendix}\label{Appendix}
\begin{figure*}
\centering
\includegraphics[width=6 in]{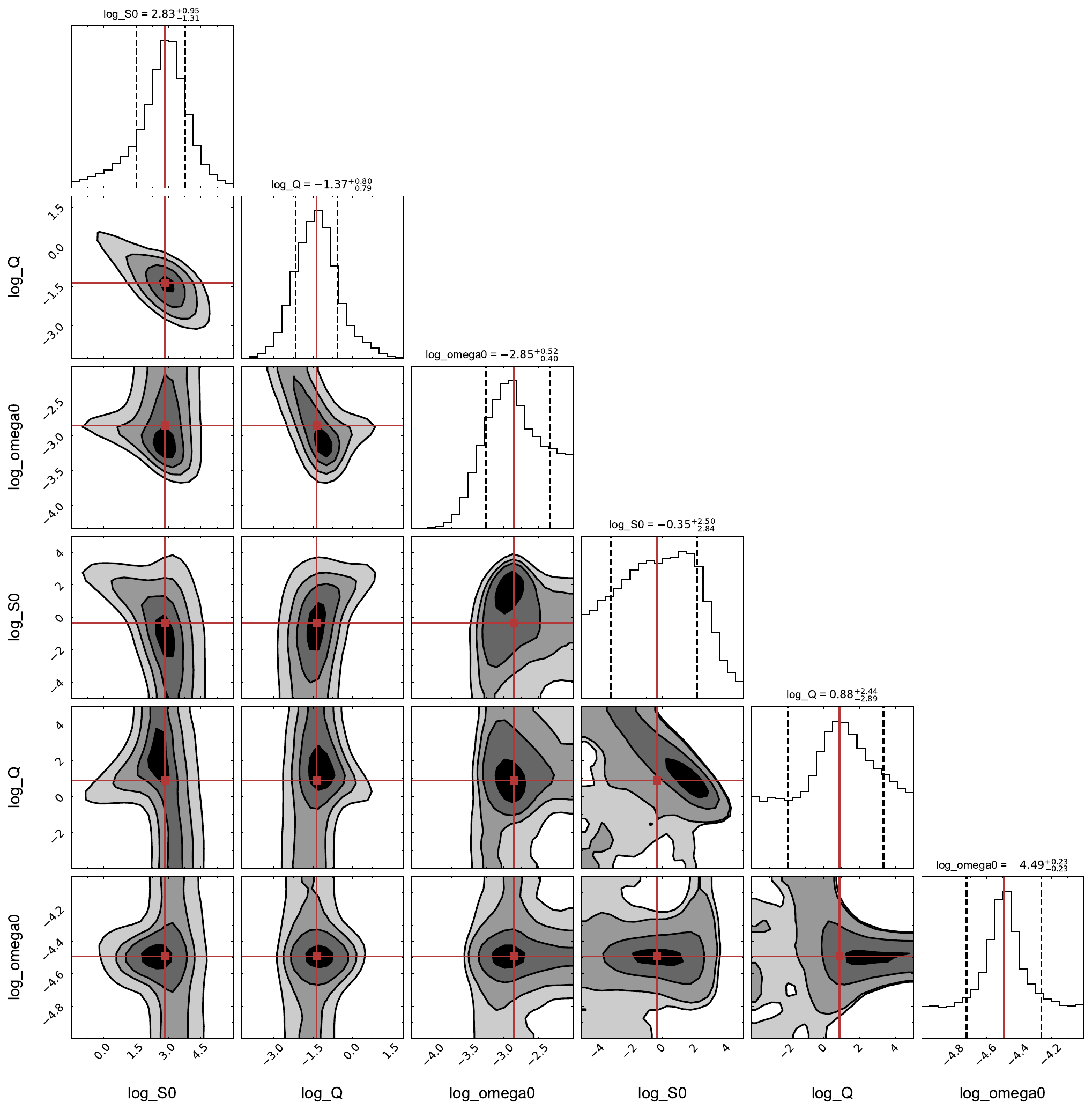}
\caption{
Posterior distributions from the MCMC analysis for the Gaussian Process with SHO$\times$2 model.
The best-fit values are taken from the 50th percentiles, as shown by the red solid lines, and the associated uncertainties correspond to the 16th and 84th percentiles. 
}
\label{fig:Cornor_GP}
\end{figure*}

The posterior distributions of parameters for the Gaussian Process are shown in Figure \ref{fig:Cornor_GP}.

\bibliographystyle{mnras}
\bibliography{Ref} % if your bibtex file is called example.bib

\begin{thebibliography}{}
\makeatletter
\relax
\def\mn@urlcharsother{\let\do\@makeother \do\$\do\&\do\#\do\^\do\_\do\%\do\~}
\def\mn@doi{\begingroup\mn@urlcharsother \@ifnextchar [ {\mn@doi@}
  {\mn@doi@[]}}
\def\mn@doi@[#1]#2{\def\@tempa{#1}\ifx\@tempa\@empty \href
  {http://dx.doi.org/#2} {doi:#2}\else \href {http://dx.doi.org/#2} {#1}\fi
  \endgroup}
\def\mn@eprint#1#2{\mn@eprint@#1:#2::\@nil}
\def\mn@eprint@arXiv#1{\href {http://arxiv.org/abs/#1} {{\tt arXiv:#1}}}
\def\mn@eprint@dblp#1{\href {http://dblp.uni-trier.de/rec/bibtex/#1.xml}
  {dblp:#1}}
\def\mn@eprint@#1:#2:#3:#4\@nil{\def\@tempa {#1}\def\@tempb {#2}\def\@tempc
  {#3}\ifx \@tempc \@empty \let \@tempc \@tempb \let \@tempb \@tempa \fi \ifx
  \@tempb \@empty \def\@tempb {arXiv}\fi \@ifundefined
  {mn@eprint@\@tempb}{\@tempb:\@tempc}{\expandafter \expandafter \csname
  mn@eprint@\@tempb\endcsname \expandafter{\@tempc}}}

\bibitem[\protect\citeauthoryear{{Abdo} et~al.,}{{Abdo}
  et~al.}{2010}]{Abdo2010ApJ716}
{Abdo} A.~A.,  et~al., 2010, \mn@doi [\apj] {10.1088/0004-637X/716/1/30}, \href
  {https://ui.adsabs.harvard.edu/abs/2010ApJ...716...30A} {716, 30}

\bibitem[\protect\citeauthoryear{{Abdollahi} et~al.,}{{Abdollahi}
  et~al.}{2020}]{Abdollahi2020ApJS}
{Abdollahi} S.,  et~al., 2020, \mn@doi [\apjs] {10.3847/1538-4365/ab6bcb},
  \href {https://ui.adsabs.harvard.edu/abs/2020ApJS..247...33A} {247, 33}

\bibitem[\protect\citeauthoryear{{Abdollahi} et~al.,}{{Abdollahi}
  et~al.}{2022}]{Abdollahi2022ApJS}
{Abdollahi} S.,  et~al., 2022, \mn@doi [\apjs] {10.3847/1538-4365/ac6751},
  \href {https://ui.adsabs.harvard.edu/abs/2022ApJS..260...53A} {260, 53}

\bibitem[\protect\citeauthoryear{{Abraham} \& {Carrara}}{{Abraham} \&
  {Carrara}}{1998}]{Abraham1998ApJ}
{Abraham} Z.,  {Carrara} E.~A.,  1998, \mn@doi [\apj] {10.1086/305387}, \href
  {https://ui.adsabs.harvard.edu/abs/1998ApJ...496..172A} {496, 172}

\bibitem[\protect\citeauthoryear{{Abraham} \& {Romero}}{{Abraham} \&
  {Romero}}{1999}]{Abraham1999A&A}
{Abraham} Z.,  {Romero} G.~E.,  1999, \aap, \href
  {https://ui.adsabs.harvard.edu/abs/1999A&A...344...61A} {344, 61}

\bibitem[\protect\citeauthoryear{{Accadia} et~al.,}{{Accadia}
  et~al.}{2012}]{Accadia2012JInst}
{Accadia} T.,  et~al., 2012, \mn@doi [Journal of Instrumentation]
  {10.1088/1748-0221/7/03/P03012}, \href
  {https://ui.adsabs.harvard.edu/abs/2012JInst...7.3012A} {7, 3012}

\bibitem[\protect\citeauthoryear{{Ackermann} et~al.,}{{Ackermann}
  et~al.}{2015}]{Ackermann2015ApJ}
{Ackermann} M.,  et~al., 2015, \mn@doi [\apjl] {10.1088/2041-8205/813/2/L41},
  \href {https://ui.adsabs.harvard.edu/abs/2015ApJ...813L..41A} {813, L41}

\bibitem[\protect\citeauthoryear{{Agazie} et~al.,}{{Agazie}
  et~al.}{2023a}]{Agazie2023ApJ951L9A}
{Agazie} G.,  et~al., 2023a, \mn@doi [\apjl] {10.3847/2041-8213/acda9a}, \href
  {https://ui.adsabs.harvard.edu/abs/2023ApJ...951L...9A} {951, L9}

\bibitem[\protect\citeauthoryear{{Agazie} et~al.,}{{Agazie}
  et~al.}{2023b}]{Agazie2023ApJ951L10A}
{Agazie} G.,  et~al., 2023b, \mn@doi [\apjl] {10.3847/2041-8213/acda88}, \href
  {https://ui.adsabs.harvard.edu/abs/2023ApJ...951L..10A} {951, L10}

\bibitem[\protect\citeauthoryear{{Atwood} et~al.,}{{Atwood}
  et~al.}{2009}]{Atwood2009apj}
{Atwood} W.~B.,  et~al., 2009, \mn@doi [\apj] {10.1088/0004-637X/697/2/1071},
  \href {https://ui.adsabs.harvard.edu/abs/2009ApJ...697.1071A} {697, 1071}

\bibitem[\protect\citeauthoryear{{Baluev}}{{Baluev}}{2008}]{Baluev2008}
{Baluev} R.~V.,  2008, \mn@doi [\mnras] {10.1111/j.1365-2966.2008.12689.x},
  \href {https://ui.adsabs.harvard.edu/abs/2008MNRAS.385.1279B} {385, 1279}

\bibitem[\protect\citeauthoryear{{Banerjee}, {Sharma}, {Mandal}, {Das},
  {Bhatta}  \& {Bose}}{{Banerjee} et~al.}{2023}]{Banerjee2023MNRAS}
{Banerjee} A.,  {Sharma} A.,  {Mandal} A.,  {Das} A.~K.,  {Bhatta} G.,   {Bose}
  D.,  2023, \mn@doi [\mnras] {10.1093/mnrasl/slad057}, \href
  {https://ui.adsabs.harvard.edu/abs/2023MNRAS.523L..52B} {523, L52}

\bibitem[\protect\citeauthoryear{{Bassa}, {Wang}, {Cumming}  \&
  {Kaspi}}{{Bassa} et~al.}{2008}]{Manchester2008AIPC}
{Bassa} C.,  {Wang} Z.,  {Cumming} A.,   {Kaspi} V.~M.,  eds, 2008, {40 YEARS
  OF PULSARS: Millisecond Pulsars, Magnetars and More}  American Institute of
  Physics Conference Series Vol. 983.
AIP

\bibitem[\protect\citeauthoryear{{Bell} et~al.,}{{Bell}
  et~al.}{2011}]{Bell2011MNRAS411}
{Bell} M.~E.,  et~al., 2011, \mn@doi [\mnras]
  {10.1111/j.1365-2966.2010.17692.x}, \href
  {https://ui.adsabs.harvard.edu/abs/2011MNRAS.411..402B} {411, 402}

\bibitem[\protect\citeauthoryear{{Bhatta}}{{Bhatta}}{2019}]{Bhatta2019MNRAS}
{Bhatta} G.,  2019, \mn@doi [\mnras] {10.1093/mnras/stz1482}, \href
  {https://ui.adsabs.harvard.edu/abs/2019MNRAS.487.3990B} {487, 3990}

\bibitem[\protect\citeauthoryear{{Bhatta} \& {Dhital}}{{Bhatta} \&
  {Dhital}}{2020}]{Bhatta2020ApJ891}
{Bhatta} G.,  {Dhital} N.,  2020, \mn@doi [\apj] {10.3847/1538-4357/ab7455},
  \href {https://ui.adsabs.harvard.edu/abs/2020ApJ...891..120B} {891, 120}

\bibitem[\protect\citeauthoryear{{Blandford} \& {Koenigl}}{{Blandford} \&
  {Koenigl}}{1979}]{Blandford1979}
{Blandford} R.~D.,  {Koenigl} A.,  1979, \aplett, \href
  {https://ui.adsabs.harvard.edu/abs/1979ApL....20...15B} {20, 15}

\bibitem[\protect\citeauthoryear{{Britzen} et~al.,}{{Britzen}
  et~al.}{2018}]{Britzen2018MNRAS}
{Britzen} S.,  et~al., 2018, \mn@doi [\mnras] {10.1093/mnras/sty1026}, \href
  {https://ui.adsabs.harvard.edu/abs/2018MNRAS.478.3199B} {478, 3199}

\bibitem[\protect\citeauthoryear{{Camenzind} \& {Krockenberger}}{{Camenzind} \&
  {Krockenberger}}{1992}]{Camenzind1992A&A}
{Camenzind} M.,  {Krockenberger} M.,  1992, \aap, \href
  {https://ui.adsabs.harvard.edu/abs/1992A&A...255...59C} {255, 59}

\bibitem[\protect\citeauthoryear{{Celotti}, {Padovani}  \&
  {Ghisellini}}{{Celotti} et~al.}{1997}]{Celotti1997}
{Celotti} A.,  {Padovani} P.,   {Ghisellini} G.,  1997, \mn@doi [\mnras]
  {10.1093/mnras/286.2.415}, \href
  {https://ui.adsabs.harvard.edu/abs/1997MNRAS.286..415C} {286, 415}

\bibitem[\protect\citeauthoryear{{Chakrabarti} \& {Wiita}}{{Chakrabarti} \&
  {Wiita}}{1993}]{Chakrabarti1993ApJ}
{Chakrabarti} S.~K.,  {Wiita} P.~J.,  1993, \mn@doi [\apj] {10.1086/172862},
  \href {https://ui.adsabs.harvard.edu/abs/1993ApJ...411..602C} {411, 602}

\bibitem[\protect\citeauthoryear{{Chalumeau} et~al.,}{{Chalumeau}
  et~al.}{2022}]{Chalumeau2022MNRAS}
{Chalumeau} A.,  et~al., 2022, \mn@doi [\mnras] {10.1093/mnras/stab3283}, \href
  {https://ui.adsabs.harvard.edu/abs/2022MNRAS.509.5538C} {509, 5538}

\bibitem[\protect\citeauthoryear{{Chatterjee} et~al.,}{{Chatterjee}
  et~al.}{2008}]{Chatterjee2008ApJ689}
{Chatterjee} R.,  et~al., 2008, \mn@doi [\apj] {10.1086/592598}, \href
  {https://ui.adsabs.harvard.edu/abs/2008ApJ...689...79C} {689, 79}

\bibitem[\protect\citeauthoryear{{Chen}, {Yu}, {Huang}  \& {Ding}}{{Chen}
  et~al.}{2024}]{Chen2024MNRAS}
{Chen} J.,  {Yu} J.,  {Huang} W.,   {Ding} N.,  2024, \mn@doi [\mnras]
  {10.1093/mnras/stae416}, \href
  {https://ui.adsabs.harvard.edu/abs/2024MNRAS.528.6807C} {528, 6807}

\bibitem[\protect\citeauthoryear{{Connolly}}{{Connolly}}{2016}]{Connolly2016ascl}
{Connolly} S.~D.,  2016, {DELightcurveSimulation: Light curve simulation code},
  Astrophysics Source Code Library, record ascl:1602.012

\bibitem[\protect\citeauthoryear{{Dong}, {Gai}, {Tang}, {Wang}  \& {Yi}}{{Dong}
  et~al.}{2022}]{Dong2022RAA}
{Dong} F.-T.,  {Gai} N.,  {Tang} Y.,  {Wang} Y.-F.,   {Yi} T.-F.,  2022,
  \mn@doi [Research in Astronomy and Astrophysics] {10.1088/1674-4527/ac71fc},
  \href {https://ui.adsabs.harvard.edu/abs/2022RAA....22k5001D} {22, 115001}

\bibitem[\protect\citeauthoryear{{Edelson} \& {Krolik}}{{Edelson} \&
  {Krolik}}{1988}]{Edelson1988APJ333}
{Edelson} R.~A.,  {Krolik} J.~H.,  1988, \mn@doi [\apj] {10.1086/166773}, \href
  {https://ui.adsabs.harvard.edu/abs/1988ApJ...333..646E} {333, 646}

\bibitem[\protect\citeauthoryear{{Emmanoulopoulos}, {McHardy}  \&
  {Papadakis}}{{Emmanoulopoulos} et~al.}{2013}]{Emmanoulopoulos2013}
{Emmanoulopoulos} D.,  {McHardy} I.~M.,   {Papadakis} I.~E.,  2013, \mn@doi
  [\mnras] {10.1093/mnras/stt764}, \href
  {https://ui.adsabs.harvard.edu/abs/2013MNRAS.433..907E} {433, 907}

\bibitem[\protect\citeauthoryear{{Fan}, {Huang}, {He}, {Yang}, {Hua}, {Liu}  \&
  {Wang}}{{Fan} et~al.}{2009}]{Fan2009PASJ}
{Fan} J.-H.,  {Huang} Y.,  {He} T.-M.,  {Yang} J.~H.,  {Hua} T.~X.,  {Liu} Y.,
   {Wang} Y.~X.,  2009, \mn@doi [\pasj] {10.1093/pasj/61.4.639}, \href
  {https://ui.adsabs.harvard.edu/abs/2009PASJ...61..639F} {61, 639}

\bibitem[\protect\citeauthoryear{{Fan}, {Liu}, {Qian}, {Tao}, {Shen}, {Zhang},
  {Huang}  \& {Wang}}{{Fan} et~al.}{2010}]{Fan2010RAA}
{Fan} J.-H.,  {Liu} Y.,  {Qian} B.-C.,  {Tao} J.,  {Shen} Z.-Q.,  {Zhang}
  J.-S.,  {Huang} Y.,   {Wang} J.,  2010, \mn@doi [Research in Astronomy and
  Astrophysics] {10.1088/1674-4527/10/11/002}, \href
  {https://ui.adsabs.harvard.edu/abs/2010RAA....10.1100F} {10, 1100}

\bibitem[\protect\citeauthoryear{{Fan}, {Kurtanidze}, {Liu}, {Richter},
  {Chanishvili}  \& {Yuan}}{{Fan} et~al.}{2014}]{Fan2014ApJS}
{Fan} J.~H.,  {Kurtanidze} O.,  {Liu} Y.,  {Richter} G.~M.,  {Chanishvili} R.,
   {Yuan} Y.~H.,  2014, \mn@doi [\apjs] {10.1088/0067-0049/213/2/26}, \href
  {https://ui.adsabs.harvard.edu/abs/2014ApJS..213...26F} {213, 26}

\bibitem[\protect\citeauthoryear{{Fan} et~al.,}{{Fan}
  et~al.}{2016}]{Fan2016ApJS}
{Fan} J.~H.,  et~al., 2016, \mn@doi [\apjs] {10.3847/0067-0049/226/2/20}, \href
  {https://ui.adsabs.harvard.edu/abs/2016ApJS..226...20F} {226, 20}

\bibitem[\protect\citeauthoryear{{Fan} et~al.,}{{Fan} et~al.}{2018}]{Fan2018AJ}
{Fan} J.~H.,  et~al., 2018, \mn@doi [\aj] {10.3847/1538-3881/aaa547}, \href
  {https://ui.adsabs.harvard.edu/abs/2018AJ....155...90F} {155, 90}

\bibitem[\protect\citeauthoryear{{Fan} et~al.,}{{Fan}
  et~al.}{2021}]{Fan2021ApJS..253...10F}
{Fan} J.~H.,  et~al., 2021, \mn@doi [\apjs] {10.3847/1538-4365/abd32d}, \href
  {https://ui.adsabs.harvard.edu/abs/2021ApJS..253...10F} {253, 10}

\bibitem[\protect\citeauthoryear{{Fermi Science Support Development
  Team}}{{Fermi Science Support Development
  Team}}{2019}]{Fermi2019asclsoft05011F}
{Fermi Science Support Development Team} 2019, {Fermitools: Fermi Science
  Tools}, Astrophysics Source Code Library, record ascl:1905.011

\bibitem[\protect\citeauthoryear{{Foreman-Mackey}, {Agol}, {Ambikasaran}  \&
  {Angus}}{{Foreman-Mackey} et~al.}{2017}]{Foreman-Mackey2017AJ154}
{Foreman-Mackey} D.,  {Agol} E.,  {Ambikasaran} S.,   {Angus} R.,  2017,
  \mn@doi [\aj] {10.3847/1538-3881/aa9332}, \href
  {https://ui.adsabs.harvard.edu/abs/2017AJ....154..220F} {154, 220}

\bibitem[\protect\citeauthoryear{{Foster}}{{Foster}}{1996}]{Foster1996}
{Foster} G.,  1996, \mn@doi [\aj] {10.1086/118137}, \href
  {https://ui.adsabs.harvard.edu/abs/1996AJ....112.1709F} {112, 1709}

\bibitem[\protect\citeauthoryear{{Ghisellini} \& {Tavecchio}}{{Ghisellini} \&
  {Tavecchio}}{2008}]{Ghisellini2008MNRAS387}
{Ghisellini} G.,  {Tavecchio} F.,  2008, \mn@doi [\mnras]
  {10.1111/j.1365-2966.2008.13360.x}, \href
  {https://ui.adsabs.harvard.edu/abs/2008MNRAS.387.1669G} {387, 1669}

\bibitem[\protect\citeauthoryear{{Ghisellini} \& {Tavecchio}}{{Ghisellini} \&
  {Tavecchio}}{2009}]{Ghisellini2009MNRAS397}
{Ghisellini} G.,  {Tavecchio} F.,  2009, \mn@doi [\mnras]
  {10.1111/j.1365-2966.2009.15007.x}, \href
  {https://ui.adsabs.harvard.edu/abs/2009MNRAS.397..985G} {397, 985}

\bibitem[\protect\citeauthoryear{{Horne} \& {Baliunas}}{{Horne} \&
  {Baliunas}}{1986}]{Horne1986ApJ}
{Horne} J.~H.,  {Baliunas} S.~L.,  1986, \mn@doi [\apj] {10.1086/164037}, \href
  {https://ui.adsabs.harvard.edu/abs/1986ApJ...302..757H} {302, 757}

\bibitem[\protect\citeauthoryear{{Hovatta}, {Valtaoja}, {Tornikoski}  \&
  {L{\"a}hteenm{\"a}ki}}{{Hovatta} et~al.}{2009}]{Hovatta2009}
{Hovatta} T.,  {Valtaoja} E.,  {Tornikoski} M.,   {L{\"a}hteenm{\"a}ki} A.,
  2009, \mn@doi [\aap] {10.1051/0004-6361:200811150}, \href
  {https://ui.adsabs.harvard.edu/abs/2009A&A...494..527H} {494, 527}

\bibitem[\protect\citeauthoryear{{Huang}, {Wang}, {Wang}  \& {Wang}}{{Huang}
  et~al.}{2013}]{Huang2013RAA}
{Huang} C.-Y.,  {Wang} D.-X.,  {Wang} J.-Z.,   {Wang} Z.-Y.,  2013, \mn@doi
  [Research in Astronomy and Astrophysics] {10.1088/1674-4527/13/6/010}, \href
  {https://ui.adsabs.harvard.edu/abs/2013RAA....13..705H} {13, 705}

\bibitem[\protect\citeauthoryear{{Jenet} et~al.,}{{Jenet}
  et~al.}{2009}]{Jenet2009arXiv0909}
{Jenet} F.,  et~al., 2009, \mn@doi [arXiv e-prints] {10.48550/arXiv.0909.1058},
  \href {https://ui.adsabs.harvard.edu/abs/2009arXiv0909.1058J} {p.
  arXiv:0909.1058}

\bibitem[\protect\citeauthoryear{{Katz}}{{Katz}}{1997}]{Katz1997APJ}
{Katz} J.~I.,  1997, \mn@doi [\apj] {10.1086/303811}, \href
  {https://ui.adsabs.harvard.edu/abs/1997ApJ...478..527K} {478, 527}

\bibitem[\protect\citeauthoryear{{LIGO Scientific Collaboration} et~al.,}{{LIGO
  Scientific Collaboration} et~al.}{2015}]{LIGO2015CQGra..32g4001L}
{LIGO Scientific Collaboration} et~al., 2015, \mn@doi [Classical and Quantum
  Gravity] {10.1088/0264-9381/32/7/074001}, \href
  {https://ui.adsabs.harvard.edu/abs/2015CQGra..32g4001L} {32, 074001}

\bibitem[\protect\citeauthoryear{{Lee}}{{Lee}}{2016}]{Lee2016ASPC..502...19L}
{Lee} K.~J.,  2016, in {Qain} L.,  {Li} D.,  eds,  Astronomical Society of the
  Pacific Conference Series Vol. 502, Frontiers in Radio Astronomy and FAST
  Early Sciences Symposium 2015. p.~19

\bibitem[\protect\citeauthoryear{{Lehto} \& {Valtonen}}{{Lehto} \&
  {Valtonen}}{1996}]{Lehto1996ApJ}
{Lehto} H.~J.,  {Valtonen} M.~J.,  1996, \mn@doi [\apj] {10.1086/176962}, \href
  {https://ui.adsabs.harvard.edu/abs/1996ApJ...460..207L} {460, 207}

\bibitem[\protect\citeauthoryear{{Lense} \& {Thirring}}{{Lense} \&
  {Thirring}}{1918}]{Lense1918PhyZ}
{Lense} J.,  {Thirring} H.,  1918, Physikalische Zeitschrift, \href
  {https://ui.adsabs.harvard.edu/abs/1918PhyZ...19..156L} {19, 156}

\bibitem[\protect\citeauthoryear{{Li}, {Luo}, {Yang}, {Yang}, {Cai}  \&
  {Yang}}{{Li} et~al.}{2017}]{Li2017ApJ}
{Li} X.-P.,  {Luo} Y.-H.,  {Yang} H.-Y.,  {Yang} C.,  {Cai} Y.,   {Yang} H.-T.,
   2017, \mn@doi [\apj] {10.3847/1538-4357/aa86ee}, \href
  {https://ui.adsabs.harvard.edu/abs/2017ApJ...847....8L} {847, 8}

\bibitem[\protect\citeauthoryear{{Li}, {Gao}, {Qin}, {Yi}  \& {Chen}}{{Li}
  et~al.}{2022}]{Li2022RAA22}
{Li} H.-Z.,  {Gao} Q.-G.,  {Qin} L.-H.,  {Yi} T.-F.,   {Chen} Q.-R.,  2022,
  \mn@doi [Research in Astronomy and Astrophysics] {10.1088/1674-4527/ac630e},
  \href {https://ui.adsabs.harvard.edu/abs/2022RAA....22e5017L} {22, 055017}

\bibitem[\protect\citeauthoryear{{Li}, {Yang}, {Cai}, {Song}, {Yang}  \&
  {Shan}}{{Li} et~al.}{2023}]{Li2023RAA23}
{Li} X.-P.,  {Yang} H.-Y.,  {Cai} Y.,  {Song} X.-F.,  {Yang} H.-T.,   {Shan}
  Y.-Q.,  2023, \mn@doi [Research in Astronomy and Astrophysics]
  {10.1088/1674-4527/ace091}, \href
  {https://ui.adsabs.harvard.edu/abs/2023RAA....23i5010L} {23, 095010}

\bibitem[\protect\citeauthoryear{{Liodakis}, {Hovatta}, {Huppenkothen},
  {Kiehlmann}, {Max-Moerbeck}  \& {Readhead}}{{Liodakis}
  et~al.}{2018}]{Liodakis2018}
{Liodakis} I.,  {Hovatta} T.,  {Huppenkothen} D.,  {Kiehlmann} S.,
  {Max-Moerbeck} W.,   {Readhead} A. C.~S.,  2018, \mn@doi [\apj]
  {10.3847/1538-4357/aae2b7}, \href
  {https://ui.adsabs.harvard.edu/abs/2018ApJ...866..137L} {866, 137}

\bibitem[\protect\citeauthoryear{{Lomb}}{{Lomb}}{1976}]{Lomb1976}
{Lomb} N.~R.,  1976, \mn@doi [\apss] {10.1007/BF00648343}, \href
  {https://ui.adsabs.harvard.edu/abs/1976Ap&SS..39..447L} {39, 447}

\bibitem[\protect\citeauthoryear{{Lynden-Bell}}{{Lynden-Bell}}{1969}]{Lynden-Bell1969}
{Lynden-Bell} D.,  1969, \mn@doi [\nat] {10.1038/223690a0}, \href
  {https://ui.adsabs.harvard.edu/abs/1969Natur.223..690L} {223, 690}

\bibitem[\protect\citeauthoryear{Maggiore}{Maggiore}{2007}]{maggiore_gravitational_2007}
Maggiore M.,  2007, Gravitational {Waves}: {Volume} 1: {Theory} and
  {Experiments}.
Oxford University Press, \mn@doi{10.1093/acprof:oso/9780198570745.001.0001},
  \url {https://doi.org/10.1093/acprof:oso/9780198570745.001.0001}

\bibitem[\protect\citeauthoryear{{Manchester} \& {IPTA}}{{Manchester} \&
  {IPTA}}{2013}]{Manchester2013CQGra..30v4010M}
{Manchester} R.~N.,  {IPTA} 2013, \mn@doi [Classical and Quantum Gravity]
  {10.1088/0264-9381/30/22/224010}, \href
  {https://ui.adsabs.harvard.edu/abs/2013CQGra..30v4010M} {30, 224010}

\bibitem[\protect\citeauthoryear{{Mangalam} \& {Wiita}}{{Mangalam} \&
  {Wiita}}{1993}]{Mangalam1993ApJ}
{Mangalam} A.~V.,  {Wiita} P.~J.,  1993, \mn@doi [\apj] {10.1086/172453}, \href
  {https://ui.adsabs.harvard.edu/abs/1993ApJ...406..420M} {406, 420}

\bibitem[\protect\citeauthoryear{{Max-Moerbeck}, {Richards}, {Hovatta},
  {Pavlidou}, {Pearson}  \& {Readhead}}{{Max-Moerbeck}
  et~al.}{2014}]{Max-Moerbeck2014MNRAS445}
{Max-Moerbeck} W.,  {Richards} J.~L.,  {Hovatta} T.,  {Pavlidou} V.,  {Pearson}
  T.~J.,   {Readhead} A.~C.~S.,  2014, \mn@doi [\mnras]
  {10.1093/mnras/stu1707}, \href
  {https://ui.adsabs.harvard.edu/abs/2014MNRAS.445..437M} {445, 437}

\bibitem[\protect\citeauthoryear{{M{\"u}cke} \& {Protheroe}}{{M{\"u}cke} \&
  {Protheroe}}{2001}]{Mucke2001}
{M{\"u}cke} A.,  {Protheroe} R.~J.,  2001, \mn@doi [Astroparticle Physics]
  {10.1016/S0927-6505(00)00141-9}, \href
  {https://ui.adsabs.harvard.edu/abs/2001APh....15..121M} {15, 121}

\bibitem[\protect\citeauthoryear{{O'Neill} et~al.,}{{O'Neill}
  et~al.}{2022}]{O'Neill2022ApJ926L}
{O'Neill} S.,  et~al., 2022, \mn@doi [\apjl] {10.3847/2041-8213/ac504b}, \href
  {https://ui.adsabs.harvard.edu/abs/2022ApJ...926L..35O} {926, L35}

\bibitem[\protect\citeauthoryear{{Otero-Santos} et~al.,}{{Otero-Santos}
  et~al.}{2020}]{Otero-Santos2020MNRAS}
{Otero-Santos} J.,  et~al., 2020, \mn@doi [\mnras] {10.1093/mnras/staa134},
  \href {https://ui.adsabs.harvard.edu/abs/2020MNRAS.492.5524O} {492, 5524}

\bibitem[\protect\citeauthoryear{{Ouyang} et~al.,}{{Ouyang}
  et~al.}{2025}]{Ouyang2025ApJ980}
{Ouyang} Z.,  et~al., 2025, \mn@doi [\apj] {10.3847/1538-4357/ada3bc}, \href
  {https://ui.adsabs.harvard.edu/abs/2025ApJ...980...19O} {980, 19}

\bibitem[\protect\citeauthoryear{{Padovani}}{{Padovani}}{2017}]{Padovani2017FrASS}
{Padovani} P.,  2017, \mn@doi [Frontiers in Astronomy and Space Sciences]
  {10.3389/fspas.2017.00035}, \href
  {https://ui.adsabs.harvard.edu/abs/2017FrASS...4...35P} {4, 35}

\bibitem[\protect\citeauthoryear{{Planck Collaboration} et~al.,}{{Planck
  Collaboration} et~al.}{2020}]{Planck2020A&A}
{Planck Collaboration} et~al., 2020, \mn@doi [\aap]
  {10.1051/0004-6361/201833910}, \href
  {https://ui.adsabs.harvard.edu/abs/2020A&A...641A...6P} {641, A6}

\bibitem[\protect\citeauthoryear{{Prokhorov} \& {Moraghan}}{{Prokhorov} \&
  {Moraghan}}{2017}]{Prokhorov2017MNRAS}
{Prokhorov} D.~A.,  {Moraghan} A.,  2017, \mn@doi [\mnras]
  {10.1093/mnras/stx1742}, \href
  {https://ui.adsabs.harvard.edu/abs/2017MNRAS.471.3036P} {471, 3036}

\bibitem[\protect\citeauthoryear{{Qian} et~al.,}{{Qian}
  et~al.}{2007}]{Qian2007ChJAA}
{Qian} S.-J.,  et~al., 2007, \mn@doi [\cjaa] {10.1088/1009-9271/7/3/05}, \href
  {https://ui.adsabs.harvard.edu/abs/2007ChJAA...7..364Q} {7, 364}

\bibitem[\protect\citeauthoryear{{Rani}, {Wiita}  \& {Gupta}}{{Rani}
  et~al.}{2009}]{Rani2009ApJ}
{Rani} B.,  {Wiita} P.~J.,   {Gupta} A.~C.,  2009, \mn@doi [\apj]
  {10.1088/0004-637X/696/2/2170}, \href
  {https://ui.adsabs.harvard.edu/abs/2009ApJ...696.2170R} {696, 2170}

\bibitem[\protect\citeauthoryear{{Rieger}}{{Rieger}}{2004}]{Rieger2004ApJ}
{Rieger} F.~M.,  2004, \mn@doi [\apjl] {10.1086/426018}, \href
  {https://ui.adsabs.harvard.edu/abs/2004ApJ...615L...5R} {615, L5}

\bibitem[\protect\citeauthoryear{{Salpeter}}{{Salpeter}}{1964}]{Salpeter1964}
{Salpeter} E.~E.,  1964, \mn@doi [\apj] {10.1086/147973}, \href
  {https://ui.adsabs.harvard.edu/abs/1964ApJ...140..796S} {140, 796}

\bibitem[\protect\citeauthoryear{{Sandrinelli}, {Covino}, {Dotti}  \&
  {Treves}}{{Sandrinelli} et~al.}{2016}]{Sandrinelli2016}
{Sandrinelli} A.,  {Covino} S.,  {Dotti} M.,   {Treves} A.,  2016, \mn@doi
  [\aj] {10.3847/0004-6256/151/3/54}, \href
  {https://ui.adsabs.harvard.edu/abs/2016AJ....151...54S} {151, 54}

\bibitem[\protect\citeauthoryear{{Scargle}}{{Scargle}}{1982}]{Scargle1982}
{Scargle} J.~D.,  1982, \mn@doi [\apj] {10.1086/160554}, \href
  {https://ui.adsabs.harvard.edu/abs/1982ApJ...263..835S} {263, 835}

\bibitem[\protect\citeauthoryear{{Schramm} et~al.,}{{Schramm}
  et~al.}{1993}]{Schramm1993A&A}
{Schramm} K.~J.,  et~al., 1993, \aap, \href
  {https://ui.adsabs.harvard.edu/abs/1993A&A...278..391S} {278, 391}

\bibitem[\protect\citeauthoryear{{Schulz} \& {Mudelsee}}{{Schulz} \&
  {Mudelsee}}{2002}]{Schulz2002CG}
{Schulz} M.,  {Mudelsee} M.,  2002, \mn@doi [Computers and Geosciences]
  {10.1016/S0098-3004(01)00044-9}, \href
  {https://ui.adsabs.harvard.edu/abs/2002CG.....28..421S} {28, 421}

\bibitem[\protect\citeauthoryear{{Sesana}}{{Sesana}}{2013}]{Sesana2013MNRAS}
{Sesana} A.,  2013, \mn@doi [\mnras] {10.1093/mnrasl/slt034}, \href
  {https://ui.adsabs.harvard.edu/abs/2013MNRAS.433L...1S} {433, L1}

\bibitem[\protect\citeauthoryear{Shapiro \& Wilk}{Shapiro \&
  Wilk}{1965}]{Shapro1965Biometrika}
Shapiro S.~S.,  Wilk M.~B.,  1965, \mn@doi [Biometrika]
  {10.1093/biomet/52.3-4.591}, 52, 591

\bibitem[\protect\citeauthoryear{{Shaw} et~al.,}{{Shaw}
  et~al.}{2012}]{Shaw2012ApJ}
{Shaw} M.~S.,  et~al., 2012, \mn@doi [\apj] {10.1088/0004-637X/748/1/49}, \href
  {https://ui.adsabs.harvard.edu/abs/2012ApJ...748...49S} {748, 49}

\bibitem[\protect\citeauthoryear{{Sillanpaa}, {Haarala}, {Valtonen},
  {Sundelius}  \& {Byrd}}{{Sillanpaa} et~al.}{1988}]{Sillanpa1988ApJ}
{Sillanpaa} A.,  {Haarala} S.,  {Valtonen} M.~J.,  {Sundelius} B.,   {Byrd}
  G.~G.,  1988, \mn@doi [\apj] {10.1086/166033}, \href
  {https://ui.adsabs.harvard.edu/abs/1988ApJ...325..628S} {325, 628}

\bibitem[\protect\citeauthoryear{{Smith} et~al.,}{{Smith}
  et~al.}{2020}]{Smith2020PASP132}
{Smith} K.~W.,  et~al., 2020, \mn@doi [\pasp] {10.1088/1538-3873/ab936e}, \href
  {https://ui.adsabs.harvard.edu/abs/2020PASP..132h5002S} {132, 085002}

\bibitem[\protect\citeauthoryear{{Spiewak} et~al.,}{{Spiewak}
  et~al.}{2022}]{Spiewak2022PASA...39...27S}
{Spiewak} R.,  et~al., 2022, \mn@doi [\pasa] {10.1017/pasa.2022.19}, \href
  {https://ui.adsabs.harvard.edu/abs/2022PASA...39...27S} {39, e027}

\bibitem[\protect\citeauthoryear{{Timmer} \& {K{\"o}nig}}{{Timmer} \&
  {K{\"o}nig}}{1995}]{Timmer1995}
{Timmer} J.,  {K{\"o}nig} M.,  1995, \aap, \href
  {https://ui.adsabs.harvard.edu/abs/1995A&A...300..707T} {300, 707}

\bibitem[\protect\citeauthoryear{{Tonry} et~al.,}{{Tonry}
  et~al.}{2018a}]{Tonry2018PASP130}
{Tonry} J.~L.,  et~al., 2018a, \mn@doi [\pasp] {10.1088/1538-3873/aabadf},
  \href {https://ui.adsabs.harvard.edu/abs/2018PASP..130f4505T} {130, 064505}

\bibitem[\protect\citeauthoryear{{Tonry} et~al.,}{{Tonry}
  et~al.}{2018b}]{Tonry2018ApJ867}
{Tonry} J.~L.,  et~al., 2018b, \mn@doi [\apj] {10.3847/1538-4357/aae386}, \href
  {https://ui.adsabs.harvard.edu/abs/2018ApJ...867..105T} {867, 105}

\bibitem[\protect\citeauthoryear{{Urry} \& {Padovani}}{{Urry} \&
  {Padovani}}{1995}]{Urry1995PASP}
{Urry} C.~M.,  {Padovani} P.,  1995, \mn@doi [\pasp] {10.1086/133630}, \href
  {https://ui.adsabs.harvard.edu/abs/1995PASP..107..803U} {107, 803}

\bibitem[\protect\citeauthoryear{{Uttley}, {McHardy}  \& {Papadakis}}{{Uttley}
  et~al.}{2002}]{Uttley2002}
{Uttley} P.,  {McHardy} I.~M.,   {Papadakis} I.~E.,  2002, \mn@doi [\mnras]
  {10.1046/j.1365-8711.2002.05298.x}, \href
  {https://ui.adsabs.harvard.edu/abs/2002MNRAS.332..231U} {332, 231}

\bibitem[\protect\citeauthoryear{{Uttley}, {McHardy}  \& {Vaughan}}{{Uttley}
  et~al.}{2005}]{Uttley2005MNRAS359}
{Uttley} P.,  {McHardy} I.~M.,   {Vaughan} S.,  2005, \mn@doi [\mnras]
  {10.1111/j.1365-2966.2005.08886.x}, \href
  {https://ui.adsabs.harvard.edu/abs/2005MNRAS.359..345U} {359, 345}

\bibitem[\protect\citeauthoryear{{Valtonen} et~al.,}{{Valtonen}
  et~al.}{2006}]{Valtonen2006ApJ}
{Valtonen} M.~J.,  et~al., 2006, \mn@doi [\apj] {10.1086/504884}, \href
  {https://ui.adsabs.harvard.edu/abs/2006ApJ...646...36V} {646, 36}

\bibitem[\protect\citeauthoryear{{Villata}, {Raiteri}, {Sillanpaa}  \&
  {Takalo}}{{Villata} et~al.}{1998}]{Villata1998MNRAS}
{Villata} M.,  {Raiteri} C.~M.,  {Sillanpaa} A.,   {Takalo} L.~O.,  1998,
  \mn@doi [\mnras] {10.1046/j.1365-8711.1998.01244.x}, \href
  {https://ui.adsabs.harvard.edu/abs/1998MNRAS.293L..13V} {293, L13}

\bibitem[\protect\citeauthoryear{{Wang}, {Cai}  \& {Fan}}{{Wang}
  et~al.}{2022}]{Wang2022ApJ}
{Wang} G.~G.,  {Cai} J.~T.,   {Fan} J.~H.,  2022, \mn@doi [\apj]
  {10.3847/1538-4357/ac5b08}, \href
  {https://ui.adsabs.harvard.edu/abs/2022ApJ...929..130W} {929, 130}

\bibitem[\protect\citeauthoryear{{Xiao} et~al.,}{{Xiao}
  et~al.}{2024}]{Xiao2024RAA24}
{Xiao} H.-B.,  et~al., 2024, \mn@doi [Research in Astronomy and Astrophysics]
  {10.1088/1674-4527/ad420e}, \href
  {https://ui.adsabs.harvard.edu/abs/2024RAA....24f5013X} {24, 065013}

\bibitem[\protect\citeauthoryear{{Xu} et~al.,}{{Xu} et~al.}{2023}]{Xu2023RAA}
{Xu} H.,  et~al., 2023, \mn@doi [Research in Astronomy and Astrophysics]
  {10.1088/1674-4527/acdfa5}, \href
  {https://ui.adsabs.harvard.edu/abs/2023RAA....23g5024X} {23, 075024}

\bibitem[\protect\citeauthoryear{{Zhang}, {Yan}, {Liao}  \& {Wang}}{{Zhang}
  et~al.}{2017}]{Zhang2017ApJ835}
{Zhang} P.-f.,  {Yan} D.-h.,  {Liao} N.-h.,   {Wang} J.-c.,  2017, \mn@doi
  [\apj] {10.3847/1538-4357/835/2/260}, \href
  {https://ui.adsabs.harvard.edu/abs/2017ApJ...835..260Z} {835, 260}

\bibitem[\protect\citeauthoryear{{Zhang}, {Yan}, {Zhang}, {Yang}  \&
  {Zhang}}{{Zhang} et~al.}{2021}]{Zhang2021ApJ}
{Zhang} H.,  {Yan} D.,  {Zhang} P.,  {Yang} S.,   {Zhang} L.,  2021, \mn@doi
  [\apj] {10.3847/1538-4357/ac0cf0}, \href
  {https://ui.adsabs.harvard.edu/abs/2021ApJ...919...58Z} {919, 58}

\bibitem[\protect\citeauthoryear{{Zhang}, {Wu}  \& {Dai}}{{Zhang}
  et~al.}{2023}]{Zhang2023PASP}
{Zhang} H.,  {Wu} F.,   {Dai} B.,  2023, \mn@doi [\pasp]
  {10.1088/1538-3873/acdf1f}, \href
  {https://ui.adsabs.harvard.edu/abs/2023PASP..135f4102Z} {135, 064102}

\bibitem[\protect\citeauthoryear{{Zhang}, {Chen}, {He}, {Nie}, {Tang}, {Huang},
  {Chen}  \& {Fan}}{{Zhang} et~al.}{2024}]{Zhang2024ApJs27127Z}
{Zhang} L.,  {Chen} X.,  {He} S.,  {Nie} W.,  {Tang} W.,  {Huang} J.,  {Chen}
  G.,   {Fan} J.,  2024, \mn@doi [\apjs] {10.3847/1538-4365/ad20c8}, \href
  {https://ui.adsabs.harvard.edu/abs/2024ApJS..271...27Z} {271, 27}

\bibitem[\protect\citeauthoryear{{Zhou}, {Wang}, {Chen}, {Wiita},
  {Vadakkumthani}, {Morrell}, {Zhang}  \& {Zhang}}{{Zhou}
  et~al.}{2018}]{Zhou2018NatCo}
{Zhou} J.,  {Wang} Z.,  {Chen} L.,  {Wiita} P.~J.,  {Vadakkumthani} J.,
  {Morrell} N.,  {Zhang} P.,   {Zhang} J.,  2018, \mn@doi [Nature
  Communications] {10.1038/s41467-018-07103-2}, \href
  {https://ui.adsabs.harvard.edu/abs/2018NatCo...9.4599Z} {9, 4599}

\makeatother
\end{thebibliography}

%\begin{thebibliography}{99}
%\end{thebibliography}

%%%%%%%%%%%%%%%%%%%%%%%%%%%%%%%%%%%%%%%%%%%%%%%%%%

%%%%%%%%%%%%%%%%% APPENDICES %%%%%%%%%%%%%%%%%%%%%

\appendix

% \section{Some extra material}

% If you want to present additional material which would interrupt the flow of the main paper,
% it can be placed in an Appendix which appears after the list of references.

%%%%%%%%%%%%%%%%%%%%%%%%%%%%%%%%%%%%%%%%%%%%%%%%%%

% Don't change these lines
\bsp	% typesetting comment
\label{lastpage}
\end{document}